\RequirePackage{fix-cm}
\documentclass[twocolumn,epjc3]{svjour3}
\smartqed  
\RequirePackage{graphicx}
\RequirePackage{mathptmx}      
\RequirePackage{latexsym}
\RequirePackage[colorlinks,citecolor=blue,urlcolor=red,linkcolor=blue]{hyperref}
\begin{document}

\title{Interacting scalar tensor cosmology in light of SNeIa, CMB, BAO and OHD observational data sets
}


\author{Sayed Wrya Rabiei\thanksref{e1,addr1}
        \and
        Haidar Sheikhahmadi\thanksref{e2,addr1,addr2}
        \and
        Khaled Saaidi\thanksref{e3,addr1,addr3}
        \and
        Ali Aghamohammadi\thanksref{e4,addr4}
}

\thankstext{e1}{e-mail: wrabiei@uok.ac.ir}
\thankstext{e2}{e-mail: h.sh.ahmadi@gmail.com/ h.sh.ahmadi@iasbs.ac.ir}
\thankstext{e3}{e-mail: ksaaidi@uok.ac.ir}
\thankstext{e4}{e-mail: a.aqamohamadi@iausdj.ac.ir}


\institute{Department of Physics, Faculty of Science, University of Kurdistan,  Sanandaj, Iran. \label{addr1}
           \and
           Institute for Advance Studies in Basic Sciences (IASBS) Gava Zang, Zanjan 45137-66731, Iran. \label{addr2}
           \and
           Science and Technology Park of Kurdistan, Sanandaj, Iran. \label{addr3}
           \and
           Sanandaj Branch Islamic Azad University, Iran. \label{addr4}
}

\date{Received: date / Accepted: date}

\maketitle

\begin{abstract}
During this work, an interacting chameleon like scalar field scenario, by considering SNeIa, CMB, BAO and OHD data sets is investigated. In fact, the investigation is realized by introducing an ansatz for the effective dark energy equation of state, which mimics the behaviour of chameleon like models. Based on this assumption, some cosmological parameters including Hubble, deceleration and coincidence parameters in such mechanism are analysed. It is realized that, to estimate the free parameters of a theoretical model, by regarding the systematic errors it better the whole of the above observational data sets to be considered. In fact, if one considers SNeIa, CMB and BAO but disregards OHD  it maybe leads to different results. Also to get a better overlap between the counters with the constraint $\chi _{\rm{m}}^2\leq 1$, the  $\chi _{\rm{T}}^2$ function could be re-weighted. The relative probability functions are plotted for marginalized likelihood $\mathcal{L} (\Omega_{\rm{m0}} ,\omega_1, \beta)$ according to two dimensional confidence levels $68.3\%$, $90\%$ and $95.4\%$. Meanwhile,  the value of free parameters which maximize the marginalized likelihoods using above confidence levels are obtained. In addition, based on these calculations the minimum value of $\chi^2$ based on free parameters of an ansatz for the effective dark energy equation of state are achieved.
\end{abstract}
\section{Introduction}
\label{sec:intro}
Observational data sets including Cosmic Microwave Background(CMB) \cite{CMB_th_1,CMB2}, Supernovae type Ia (SNeIa) \cite{ch1:7,ch1:8}, Baryonic Acoustic Oscillations (BAO) \cite{BAO_th_Eisenstein,BAO_th_Shoji}, Observational Hubble Data (OHD) \cite{FarooqRatra,H0_1}, Sloan Digital Sky Survey (SDSS) \cite{ch1:10,SD2}, and Wilkinson Microwave Anisotropy Probe (WMAP) \cite{ch1:9,WM2}, are considered as criterion for accuracy of theoretical models. Amongst these constraints, the CMB and SNeIa( because of abundance of their data sources) attract more attention. It is notable, the SNeIa constraint in high redshift values do not give a good clue to investigate the evolution of the Universe. It is obvious the results of individual observations give different values for free parameters of a theoretical model; hence, it is better that, to  estimate the best quantities for free parameters of the model  one considers the whole of observational data sets including CMB, SNeIa, BAO and OHD. Therefore, this motivated us to study the behaviour of free parameters and their overlaps. Thence,  a collective of observations including SNeIa, CMB, BAO and OHD are considered. Meanwhile the mentioned observational data sets have predicted an ambiguous form of matter which leads to an  accelerated phase of present epoch and it is well known as dark energy. Based on this ambiguous form of matter, scientists have proposed different proposals up to now. Amongst all  of those proposals, the cosmological constant, $\Lambda$, model attracts more attention \cite{Lamb1,Lamb2}. But this mechanism suffers two well known drawbacks. The first of them is related to  estimate  the contribution of quantum fluctuation of zero point energy and second is related to the ratio of $\Lambda$ and dark matter energy densities. These problems and also the excellent work by Brans and Dicke \cite{ch1:6} motivated scientists to introduce a mechanism which  $\Lambda$ had time dependency, namely quintessence \cite{Quint1,Quint2,Quint3}. Beside the quintessence mechanism, some proposals which risen from quantum gravity or string theory are introduced to estimate cosmological parameters. For instance one can refer to tachyon \cite{Tach1,Tach2}, phantom  \cite{Phant1,Phant2,Phant3}, quintum  \cite{Qui1,Qui2}, k-essense  \cite{K1,K2} and so on. Also some models which were risen from quantum field fluctuations or space time fluctuations attract more attention to investigate dark energy concept. For such models, one can mention Zero Point Quantum Fluctuations(ZPQF) \cite{ZP1,ZP2,ZP3}, Holographic Dark Energy(HDE) \cite{HD1,HD2,HDE3,HDE4,HDE5,HDE6}, Agegraphic Dark Energy (ADE) and new-ADE \cite{AGE1,Age2,Age3}. If  scalar field, in the quintessence model couples to matter (non relativistic) it induces to appear a fifth force. When the coupling is of order unity, the results of strongly coupling scalar field is not in good agreement with local gravity tests (for instance in solar system). Thus a mechanism should be exist to suppress the effect of fifth force; such mechanism is capable to reconcile strong coupling models with local experiments was proposed by Khoury and Weltman \cite{KHo1,Kho2} and also, separately, by Mota and Barrow \cite{MOF1} namely chameleon. In this mechanism, one can not choose an arbitrary Lagrangian for matter, $L_{m}$. To avoid deviation of geodesic trajectory, the author of \cite{ch1:21} has shown the best choices are $L_{m}=P$ and $L_{m}=-\rho$, where  $P$ is pressure and $\rho$ is energy density of matter, for more discussion we refer reader to \cite{GEo1,GEo2,Geo3}. Therefore main motivation of this work is  investigation the behavior of an interacting scalar field mechanism; based on these calculations and   SNeIa, CMB, BAO and OHD data sets the minimum value of $\chi^2$  for the effective dark energy equation of state are achieved.
The scheme of the paper is as follows:\\
 The above discussions which are  brief review about observational and theoretical motivations are considered as introduction. In Sec.\,2, the general theoretical discussions risen from a chameleon like mechanism related to the cosmological parameters such as Hubble, deceleration and coincidence parameters will be discussed. In Sec.\,3, a brief review about cosmological data sets are brought.  In Sec.\,4,  the observational data sets including SNeIa, CMB, BAO and OHD  are considered, to estimate the minimum value of $\chi^2$ related to free parameters of an ansatz for the effective dark energy equation of state. And  at last, Sec.\,5,  is dedicated  to concluding remarks.

\section{Conservation and field's equations in an effective dark energy scenario}
\label{sec:Conservation and field's equations}
 In chameleon like scalar field scenario, the mass of scalar field is a function of local matter density, so that, it is sufficiently large on dense environment. Due to this fact, the equivalence principal (EP) is satisfied in the laboratory {\cite{KHo1,Kho2}}. In addition, the  Brans Dicke $\omega$ parameter for two observational values of $\gamma$ post Newtonian parameter take  the values of order $10^4$  {\cite{ch19}}, which satisfies the solar system constraint.
 The  chameleon like scenario is defined as
\begin{equation}\label{eq:action1}
S=\int{d^4 x\frac{1}{2}\sqrt{-g}\Big({R-\partial ^{\mu}\varphi\,\partial_\mu\varphi-2V(\varphi)+2f(\varphi)L}\Big)},
\end{equation}
 \cite{ch1:21,GEo1,ch20,ch16,ch17,ch17a,ch17aa}. In this equation $g$ is the determinant of the metric, $V(\varphi)$ is a run away potential and the latest term indicates a non-minimal coupling between scalar field and matter sector. It should be noted $L$ is the Lagrangian density of matter which consists of both dark matter and dark energy sectors as perfect fluid \cite{ch19,ch20,ch16,ch17aa,ch18}. It should be noticed that, background is a spatially flat Friedmann-Limatire-Robertson-Walker (FLRW) Universe,  with signature $(+2)$. The variation of the action \ref{eq:action1}, with respect to (w.r.t) $g_{\mu\nu}$ results the gravitational field equation as
\begin{equation}\label{eq:Tensor_Equation_01_Metric}
G_{\mu \nu }  = f(\varphi )T_{\mu \nu }  + T_{\mu \nu }^{(\varphi )},
\end{equation}
where the stress-energy density of scalar field expresses
\begin{equation}\label{T_mu-nu}
 T_{\mu \nu }^{(\varphi )}=\left( {\nabla _\mu  \varphi \nabla _\nu  \varphi  - \frac{1}{2}g_{\mu \nu } (\nabla \varphi )^2 } \right) - g_{\mu \nu } V(\varphi ),
\end{equation}
and
\begin{equation}\label{eq:Tensor_Equation_02_Metric}
T_{\mu \nu }  = \frac{{ - 2}}{{\sqrt { - g} }}\frac{{\delta (\sqrt { - g} L)}}{{\delta g^{\mu \nu } }},
\end{equation}
is the definition of stress-energy tensor of matter. By considering $00$ and $ii$ components of $T^{(\varphi )}_{\mu \nu }$, the  energy density and pressure could be achieved. After some algebra the conservation equation reads
\begin{equation}\label{eq:Conservation_eq_01}
\nabla _\mu  (G^{\mu \nu } )= \nabla _\mu  \Big[f(\varphi )T^{\mu \nu }  + T_{(\varphi )}^{\mu \nu }\Big] = 0.
\end{equation}
In addition, the variation of the action \ref{eq:action1}, w.r.t scalar field gives the evolution equation as
\begin{equation}\label{eq:Master_phi}
\ddot \varphi  + 3H\dot \varphi  =  - V(\varphi ) + \frac{{\partial f(\varphi )}}{{\partial \varphi }}L.
\end{equation}
Now, by substituting equation \ref{eq:Master_phi} into relation \ref{eq:Conservation_eq_01}, two conservation equations for scalar field and matter are attained as
\begin{eqnarray}\label{eq:Conservation_02_phi:1}
\nabla_\mu  [T_{(\varphi )}^{\mu 0}] &=& \dot f(\varphi )L,\\
\nabla_\mu  [f(\varphi )T^{\mu 0}]& =&  - \dot f(\varphi )L, \label{eq:Conservation_02 phi:2}
\end{eqnarray}
where over $dot$ denotes derivation w.r.t ordinary cosmic time, $t$. As it was mentioned in the introduction, the Lagrangian of matter is considered as  $L= L_{(m)}  + L_{(de)}$, \cite{ch17aa,ch1:20a}; where subscript $m$ denotes matter (cold dark matter and baryons) and $de$ refers to dark energy. Then the conservation equations could be rewritten as
\begin{eqnarray}
 \nabla _\mu  \Big[f(\varphi )T_{(m)}^{\mu 0}\Big] &=&  - \dot f(\varphi )L_{(m)}, \label{eq:Conservatio 04 phi:1} \\
\nabla _\mu  \Big[f(\varphi )T_{(de)}^{\mu 0}\Big]  &=&  - \dot f(\varphi )L_{(de)}, \label{eq:Conservatio 04 phi:2}  \\
\nabla _\mu \Big[T_{(\varphi )}^{\mu 0}\Big]  &=& \dot f(\varphi )(L_{(m)}  + L_{(de)} ). \label{eq:Conservatio 04 phi:3}
\end{eqnarray}
By combining the relation \ref{eq:Tensor_Equation_01_Metric} and above equations, it is easy to receive
\begin{equation}\label{eq:Conservation_04_Tphi_Just_Lm}
\nabla _\mu  \Big[T_{(\varphi )}^{\mu 0}  + f(\varphi )T_{(de)}^{\mu 0}\Big]  = \dot f(\varphi )L_{(m)}.
\end{equation}
In the next step, by virtue of the definition of $T_{\mu \nu }^{(\varphi )}$  and equation \ref{eq:Tensor_Equation_01_Metric}, the Einstein tensor is modified as
\begin{equation}\label{eq:Modified E-tensor}
G^{\mu \nu }  = f(\varphi )\Big[ {T_{(m)}^{\mu \nu }  + T_{(de)}^{\mu \nu }  + \frac{1}{{f(\varphi )}}T_{(\varphi )}^{\mu \nu } } \Big].
\end{equation}
Hereafter, we postulate that both scalar field and dark energy, behave the same as perfect fluid, thence for such perfect mixture the effective stress-energy tensor is obtained as follows
\begin{equation}\label{eq:T_DE_effective}
T_{(DE)}^{\mu \nu }  = T_{(de)}^{\mu \nu }  + \frac{1}{{f(\varphi )}}T_{(\varphi )}^{\mu \nu },
\end{equation}
where subscript $DE$ denotes effective dark energy. Therefore using equations \ref{eq:Conservation_02_phi:1}-\ref{eq:T_DE_effective}, the modified Einstein equation and conservation relations are attained as
\begin{equation}\label{eq:Tensor_Equation_05_DE_effective}
 G^{\mu \nu }  = f(\varphi )\Big[T_{(m)}^{\mu \nu }  + T_{(DE)}^{\mu \nu }\Big],
\end{equation}
\begin{eqnarray}\label{eq:Conservation_eq_05_Tm_Lm}
 \nabla _\mu  \Big[f(\varphi )T_{(m)}^{\mu 0}\Big] &=&  - \dot f(\varphi )L_{(m)}, \label{eq:Conservation_eq_05_Tm_Lm:1}  \\
\nabla _\mu \Big[f(\varphi )T_{(DE)}^{\mu 0}\Big] &=& \dot f(\varphi )L_{(m)} . \label{eq:Conservation_eq_05_TDE_Lm:2}
\end{eqnarray}
It should be noticed that, in the right hand side of above equations, only $L_{(m)}$ is appeared. In fact it could be concluded that the energy, for different components of the Universe is not conserved separately. In refs.\cite{ch1:18,ch1:19,ch1:20}, it has been shown that, for perfect fluids, that do not couple directly
to the other components of the Universe, there are
different Lagrangian densities are equivalent. Namely, one can find that the
 two Lagrangian densities $L_{(m)} = P$ and $L_{(m)}=-\rho$ give the same stress-energy tensor and  the equation of motions for all components of the system are similar as well. But in an interacting case, which matter has an interaction with scalar field, the    Lagrangian degeneracy is broken. Based on ref.\cite{ch1:21}, the best choice for such models is $L_{(m)} = P$. Using this definition for Lagrangian of the matter one can obtain
\begin{equation}\label{eq:Cosmic_eq_06a_H2}
H^2  = \frac{1}{3}f(\varphi )\Big[\rho _m  + \rho _{DE}\Big],
\end{equation}
and also
\begin{eqnarray}\label{eq:Cosmic_eq_06b_Conservation_m}
& \frac{d}{{dt}}\Big[f(\varphi )\rho _m\Big] + 3Hf(\varphi )\rho _m  = 0, \label{eq:Cosmic_eq_06b_Conservation_m:1} \\
& \frac{d}{{dt}}\Big[f(\varphi )\rho _{DE}\Big] + 3Hf(\varphi )\Big[1 + \omega _{DE}\Big]\rho _{DE}  = 0,\label{eq:Cosmic_eq_06b_Conservation_m:2}
\end{eqnarray}
where $H=\dot{a}(t)/a(t)$ is the Hubble parameter, $a(t)$ is scale factor and $\omega _{DE}$, is the EoS parameter of the effective dark energy and satisfies EoS equation as
\begin{equation}\label{eq:P_equals_wDE_Times_rho_DE}
P_{DE}  = \omega _{DE}  \times \rho _{DE}.
\end{equation}
To establish an accurate link between theoretical results and observations, one can use the red shift parameter, $z$, instead of the scale factor; these two cosmological parameters have a relation as
\begin{eqnarray}\label{eq:Redshift_Dot_vs_H}
 \frac{{a(t_0 )}}{{a(t)}} = 1 + z \, \qquad \dot z =  - (1 + z)H.
\end{eqnarray}
Thus substituting equation \ref{eq:Redshift_Dot_vs_H} into \ref{eq:Cosmic_eq_06b_Conservation_m:1} and  \ref{eq:Cosmic_eq_06b_Conservation_m:2}, one finds out
\begin{eqnarray}\label{eq:rho_m_phi_z}\label{eq:rho_m_phi_z:1}
& f(\varphi )\rho _m  = f_0  \times \rho _{m0}  \times (1 + z)^3,  \\
& f(\varphi )\rho _{DE}  = f_0  \times \rho _{DE0}\times \exp\Big[\int_0^z {3\frac{{1 + \omega _{DE} (\tilde z)}}{{1 + \tilde z}}d\tilde z}\Big],\label{eq:rho_m_phi_z:2}
\end{eqnarray}
where $\rho _{DE0}$ and $\rho _{m0}$ refer to  energy densities of dark energy and matter at present time, respectively.
\subsection{Hubble parameter}
\label{subsec: Hubble parameter}
 Dimensionless Hubble parameter and density parameters could be defined as
\begin{equation}\label{eq:Ez_H_H0}
E(z) = \frac{H(z)}{H_0},
\end{equation}
\begin{eqnarray}\label{eq:Omega_bar_m0}
 {\bar{\Omega}} _{m0}  &=& \frac{{ \rho _{m0} }}{{3H_0^2 }}, \label{eq:Omega_bar_m0:1}\\
 {\bar{\Omega}} _{DE0}  &= &\frac{{ \rho _{DE0} }}{{3H_0^2 }}. \label{eq:Omega_bar_m0:2}
\end{eqnarray}
The dimensionless density parameters could be  rewritten as
\begin{eqnarray}\label{eq:Omega_m0}
& \Omega_ {m0}  = \frac{{f_0 \times \rho _{m0} }}{{3H_0^2 }}, \label{eq:Omega_m0:1} \\
& \Omega_ {DE0}  = \frac{{f_0 \times \rho _{DE0} }}{{3H_0^2 }}. \label{eq:Omega_m0:2}
\end{eqnarray}
Therefore using relations \ref{eq:Cosmic_eq_06a_H2} and \ref{eq:Ez_H_H0}, the dimensionless Hubble parameter is obtained as follows
\begin{eqnarray}\label{eq:E2_z}
&& E^2 (z) = \Omega _{m0} (1 + z)^3  + \Omega _{DE0} \exp \left[ {\int_0^z {3\frac{{1 + \omega _{DE} (\tilde z)}}{{1 + \tilde z}}d\tilde z} } \right].
\end{eqnarray}
\subsection{Coincidence parameter}
\label{subsec: Deceleration parameter}
The ratio  of dark matter and dark energy is defined as  coincidence parameter and could be obtained as
\begin{eqnarray} \label{eq:r_z}
 r &=& \frac{{\rho _m }}{{\rho _{DE} }} \\
 ~&=& r_0 (1 + z)^3 \exp [ - 3\int_0^z {\frac{{1 + \omega _{DE} (\tilde z)}}{{1 + \tilde z}}d\tilde z} ]. \nonumber
\end{eqnarray}
Also one can obtain
\begin{eqnarray} \label{eq:dr_dz}
&& \frac{{dr}}{{dz}}=\frac{{-3 \omega_{DE}(z)}}{{1+z}} r(z).
\end{eqnarray}
Due to the role of this parameter, $r$, in the investigation of the cosmic evolution, it attracts more attention in observational investigations. In fact one can observe that, this importance is arisen from the relation between the EoS parameter and the evolution of $r$.
\subsection{Deceleration Parameter}
\label{Deceleration Parameter}
To investigate the acceleration of the Universe, one can use deceleration parameter which is defined as
\begin{eqnarray}  \label{eq:q_t}
q(t) = \frac{{ - 1}}{{a(t)H^2 }}\frac{{d^2 a(t)}}{{dt^2 }}.
\end{eqnarray}
The above equation can be rewritten as
\begin{eqnarray}   \label{eq:q_Ez}
q(z) =  - 1 + \frac{3}{2}\left( {\frac{{(1 + \omega _{DE} )E^2  - \Omega _{m0} (1 + z)^3 \omega _{DE} }}{{(1 + z)E^2 }}} \right).
\end{eqnarray}
In present epoch of the Universe evolution, deceleration parameter is determined as
\begin{eqnarray}  \label{eq:q0_wde}
q_0  = \frac{1}{2} + \frac{3}{2}\Big[ {1 - \Omega _{m0} } \Big]\omega _{DE} (0).
\end{eqnarray}
To solve the above equation we introduce an ansatz for EoS parameter as \cite{ch1:20a,Zim}
\begin{eqnarray}  \label{eq:wDE_z_Ansatz}
\omega_{DE}(z)=-1+\omega_0 + \omega_1 (1+z)^\beta ,
\end{eqnarray}
where $\omega_0$, $\omega_1$ and $ \beta$ are  free parameters of the model, where the minimum value of $\chi^2$  of them  will be obtained in fitting part. Also it is notable if we choose $\beta=0$, the model  reduces to EoS constant models (for instance $\Lambda CDM$)~\cite{ch1:20a}.
By substituting \ref{eq:wDE_z_Ansatz} in \ref{eq:E2_z}, the dimensionless Hubble parameter  is attained as follows
\begin{eqnarray}\label{eq: E2_z_w_Ansatz}
E^2 (\{z; {\rm{P}}_{\rm{i}}\})=\Omega_{m0}(1+z)^3+
\Omega _{DE0} (1 + z)^{3\omega_0}&\times&\\ \nonumber
 \exp \left[ {3\frac{{\omega _1 }}{\beta }\left({(1 + z)^\beta- 1} \right)} \right], \label{eq:E2_z_w_Ansatz:1}
\end{eqnarray}
where
\begin{eqnarray}
\{z; {\rm{P}}_{\rm{i}}\} &=& \{\Omega _{m0} ,\omega _0 ,\omega _1 ,\beta \},\label{eq:E2_z_w_Ansatz:2}
\end{eqnarray}
and $\{\rm{P}_{\rm{i}}\}$ is a set of free parameters which should be determined using data fitting process. Using equation (\ref{eq:wDE_z_Ansatz}), one can rewrite the equations \ref{eq:rho_m_phi_z:1}, \ref{eq:rho_m_phi_z:2} and \ref{eq:r_z},  respectively as
\begin{eqnarray}\label{eq:rho_m_phi_z_Ansatz}
 f(\varphi ) \rho _m  =& f_0   \rho _{m0}\times (1 + z)^3, \label{eq:rho_m_phi_z_Ansatz:1}  \\
 f(\varphi ) \rho _{DE}  =& f_0  \rho _{DE0}\times (1 + z)^{3\omega _0 } \times\\ \nonumber
&\exp \left[ {3\frac{{\omega _1 }}{\beta }\left( {(1 + z)^\beta- 1} \right)} \right], \label{eq:rho_DE_phi_z_Ansatz:2}
\end{eqnarray}
and
\begin{equation} \label{eq:r_z_Ansatz}
r(z) = r_0 (1 + z)^{ - 3(1 - \omega _0 )} \exp \left[ { - 3\frac{{\omega _1 }}{\beta }\left( {(1 + z)^\beta   - 1} \right)} \right].
\end{equation}
\section{A brief review as to cosmological observational data sets}
\label{observational constraints}
In this section, we should emphasis that the analysis is restricted to the background level, and do not include perturbations. In the following, we want to compare our theoretical results with observations. To this end, we consider four important data sets including SNeIa, CMB, BAO and OHD. In some papers, it was claimed OHD, which obtained versus red shift, is comparable with SNeIa data set, for instance we refer reader to reference \cite{FarooqRatra} and references which are there. This subject motivated us to investigate the effects of this new data set beside other observations to improve the theoretical results. As it will be discussed, the results of OHD although is not independent of SNeIa and BAO data sets \cite{FarooqRatra}  but has not any dependency to CMB. Also there are two ways to study CMB and BAO data point among the full parameter distribution and Gaussian which in follow the latter will be used.
\subsection{Supernovae type Ia}
\label{Supernovae type Ia}
It is explicit that, supernovae attract more attention in empirical cosmology. Whereas they are very luminous, people interested to consider them, also for instance at closer distances (i.e. lower redshift) they could be used to calculate Hubble parameter, and for farther distances (i.e. higher redshift) they attain an important role to estimate deceleration parameter $q$. It is obvious there are uncertainties of different nature: statistical or random errors and systematic errors. In this work it is remarkable the systematic errors for SNeIa and OHD are neglected. In reality there is always a limit on statistical accuracy, besides the
trivial one that time for repetitions is limited. The assumption of independence is
violated in a very specific way by so-called systematic errors which appear in any
realistic experiment. For instance experiments in nuclear and particle physics usually extract the information from a statistical data sample. The precision of the results then is mainly determined by the number N of collected reactions. Besides the corresponding well defined statistical errors, nearly every measurement is subject to further uncertainties, the systematic errors, typically associated with auxiliary parameters related to the measuring apparatus, or with model assumptions. The result is typically presented in the form
\[x = 2.34 \pm 0.06 = 2.34 \pm 0.05(stat) \pm 0.03(syst).\]
The only reason for the separate quotation of the two uncertainties is that the
size of the systematic uncertainties is less well known than that of the purely statistical error \cite{Likelihood01}.
By virtue of the likelihood functions, one able to estimate the minimum value of $\chi^2$ for the set of parameters $\{{\rm{p}}_{\rm{i}}\}$, as
\begin{equation}\label{eq: likelihood}
{\cal L}(\{ {\rm{p}}_{\rm{i}} ,\mu _{\rm{0}} \} ) \propto\exp \left[ { - \frac{1}{2}\chi _{SNe}^2 (\{ {\rm{p}}_{\rm{i}} ,\mu _{\rm{0}} \} )} \right],
\end{equation}
where
\begin{equation}\label{eq: likelihood_ chi2}
\chi _{SNe}^2 (\{ {\rm{p}}_{\rm{i}} ,\mu _{\rm{0}} \} )= \sum\limits_{n= 1}^{557} {\frac{{\left[ {\mu _{obs} (z_n ) -  \mu _{th} (z_n;\rm{\{p_i , \mu_0\}}) } \right]^2 }}{{\sigma _n^2 }}}.
\end{equation}
In \ref{eq: likelihood_ chi2}, $\mu _{obs} (z_n )$ is the observational distance modulus for $n$th supernova, $\sigma _n$ is the variance of the measurement and  $\mu _{th} (z_n )$ is the theoretical distance modulus for $n$th supernova which defined as
\begin{eqnarray}\nonumber
&& \mu _{th} (z_n ;\rm{\{p_i , \mu_0\}}) = 5\log _{10} \left[ {D_L (z_n ;\rm{\{p_i\}})} \right] + \mu _0, \\\nonumber
&& \mu _0  = 42.38 - 5\log _{10} \left[ h \right], \\\nonumber
&& D_L (z_n ;\rm{\{p_i\}}) = (1 + z)\int_0^z {\frac{{d\tilde z}}{{E(\tilde z;{\rm{ \{p_i\}}})}}},
\end{eqnarray}
where $D_L$ is the luminous distance and $h=100{km s^{-1} Mpc^{-1}}$. To achieve best fit of free parameters, one can marginalize likelihood function w.r.t $\mu _{\rm{0}}$  \cite{SNeIa_Union2,Likelihood01}. Thence $\chi _{SNe}^2 (\{ {\rm{p}}_{\rm{i}}\})$  reduces to
\begin{equation}\label{eq: ABC}
\chi _{SNe}^2 (\{ {\rm{p}}_{\rm{i}}\})=A-\frac{B^2}{C},
\end{equation}
where $A$, $B$ and $C$  are defined as follows
\begin{eqnarray}\label{eq: ABC_seprated}
&& A = \sum\limits_{n = 1}^{557} {\frac{{\left[ {\mu _{obs} (z_n ) - \mu _{th} (z_n ;\{ {\rm{p}}_{\rm{i}} ,\mu _{\rm{0}}  = 0\} )} \right]^2 }}{{\sigma _n^2 }}} , \\
&& B = \sum\limits_{n = 1}^{557} {\frac{{\mu _{obs} (z_n ) - \mu _{th} (z_n ;\{ {\rm{p}}_{\rm{i}} ,\mu _{\rm{0}}  = 0\} )}}{{\sigma _n^2 }}} , \\
&& C = \sum\limits_{n = 1}^{557} {\frac{1}{{\sigma _n^2 }}}.
\end{eqnarray}
\subsection{Cosmic Microwave Background}
\label{Cosmic Microwave Background}
 According to oscillations appear in matter and radiation fields  Doppler peaks in radiation (photon) spectrum are produced. Also it should be noted that existence  of dark energy, affects the place of the Doppler peaks in spectrum diagrams. To determine the shift of these peaks, theoretically, CMB shift parameters is defined as refs. \cite{CMB_th_1,CMB_th_2}
\begin{equation}\label{eq: CMB shift parameter}
R_{th} (z_{rec} ;\{ {\rm{p}}_{\rm{i}} \} ) = \sqrt {\frac{{\Omega _{m0} }}{{f_0 }}} \int_0^{z_{rec} } {\frac{{d\tilde z}}{{E(\tilde z;\{ {\rm{p}}_{\rm{i}} \} )}}}.
\end{equation}
In CMB investigations ~\cite{CMB_obs}, the $\chi _{CMB}^2$ function versus CMB shift parameter is
\begin{equation}\label{eq: chi2 CMB}
\chi _{CMB}^2 (\{ {\rm{p}}_{\rm{i}} \} ) = \frac{{\left[ {R_{obs}  - R_{th} (z_{rec} ;\{ {\rm{p}}_{\rm{i}} \} )} \right]^2 }}{{\sigma _R^2 }}
\end{equation}
where $R_{obs} = 1.725$, $\sigma_R = 0.018$ and $z_{rec}  \approx 1091.3$ are observational quantities of CMB shift parameter, uncertainty of $R$ in $\sigma_1$ confidence level and recombination redshift, respectively refs.~\cite{CMB_th_2,CMB_th_1}.
\subsection{Baryonic Acoustic Oscillations}
\label{BAO}
 As in \cite{4} mentioned, because BAO can be considered as a standard length scale at a wide range of redshift it is an useful candidate for cosmological models testing. The importance of BAO mechanism is related to its ability in estimation the contents and curvature of the Universe. One can establish a relation between theoretical BAO parameter, $A_{th}$, and dimensionless Hubble parameter, Eq.(\ref{eq:E2_z}), as
\begin{eqnarray}\label{eq: BAO parameter}
A_{th} (z_b ;\{ {\rm{p}}_{\rm{i}} \} )=&\\ \nonumber
 & \sqrt {\frac{{\Omega _{m0} }}{{f_0 }}} \left[ {E(z_b ;\{ {\rm{p}}_{\rm{i}} \} )} \right]^{ - 1/3} \left[ {\frac{1}{{z_b }}\int_0^{z_{b} } {\frac{{d\tilde z} }{{E(\tilde z;\{ {\rm{p}}_{\rm{i}} \} )}}} } \right]^{2/3},
\end{eqnarray}
where $z_b=0.35$ ~\cite{BAO_th_Eisenstein,BAO_th_Shoji}. Also $\chi _{BAO}^2$ in BAO mechanism investigation is as follows
\begin{equation}\label{eq: BAO chi2}
\chi _{{\rm{BAO}}}^2 (\{ {\rm{p}}_{\rm{i}} \} ) = \frac{{\left[ {A_{obs}  - A_{th} (z_{rec} ;\{ {\rm{p}}_{\rm{i}} \} )} \right]^2 }}{{\sigma _A^2 }},
\end{equation}
and also $A_{\rm{obs}} =0.469(n_s / 0.98)^{-0.35}$ and $n_s = 0.968$,~\cite{BAO_th_Shoji,SNeIa_Union2}.
It is obvious that, the BAO are detected in the clustering of the combined 2dFGRS and SDSS main galaxy samples, and measure the distance-redshift relation at $z=0.2$. But we consider BAO in the clustering of the SDSS luminous red galaxies in which measure the distance-redshift relation at $z=0.35$ {\cite{Solano}}.
\subsection{Observational Hubble Data}
\label{OHD}
We suggest that, if people want to investigate  the accuracy of any theoretical model, it is better, maybe, to consider  SNeIa, CMB, BAO and OHD together. In \cite{FarooqRatra}, it was claimed that three different models of dark energy i.e. $\rm{\Lambda CDM}$, $\rm{\varphi CDM}$ and $\rm{XCDM}$ have been investigated just by considering $H(z)$ measurement.  But they have used $\bar{H}_0= 68 \pm 2.8$ and $\bar{H}_0 = 73.8\pm2.4$ which risen from SNeIa data \cite{H0_1}. Therefore it is realized that for the comparison between theoretical results and observations only OHD could not be considered. The $\chi_{OHD}^2$ function parameter based on OHD data set is defined as
\begin{equation}\label{eq: chi2 OHD}
\chi _{{\rm{OHD}}}^2 (\{ {\rm{p}}_{\rm{i}} ,H_{\rm{0}} \} ) = \sum\limits_{n = 1}^{28} {\frac{{\left[ {H_{obs} (z_n ) - H_0 E_{th} (z_n ;\{ {\rm{p}}_{\rm{i}} \} )} \right]^2 }}{{\sigma _n^2 }}},
\end{equation}
after marginalize w.r.t $H_{\rm{0}}$, to calculate likelihood, $\chi _{{\rm{OHD}}}^2$ could be considered as
\begin{equation}\label{eq: chi2_OHD _AbC}
 \chi _{{\rm{OHD}}}^2 (\{ {\rm{p}}_{\rm{i}} \} ) = A_H  - \frac{{B_H^2 }}{{C_H }},
\end{equation}
where
\begin{eqnarray}\label{eq: ABC_Chi2}
 A_H  &=& \sum\limits_{n = 1}^{28} {\frac{{\left[ {H_{obs} (z_n )} \right]^2 }}{{\sigma _n^2 }}},  \\
 B_H  &=& \sum\limits_{n = 1}^{28} {\frac{{H_{obs} (z_n ) \times E_{th} (z_n ;\{ {\rm{p}}_{\rm{i}} \} )}}{{\sigma _n^2 }}},  \\
 C_H  &=& \sum\limits_{n = 1}^{28} {\frac{{\left[ {E_{th} (z_n ;\{ {\rm{p}}_{\rm{i}} \} )} \right]^2 }}{{\sigma _n^2 }}}.
\end{eqnarray}
In above equations subscript $obs$ is refer to observational quantities and subscript $th$ is for theoretical one.

\section{Cosmological constraints and data fitting}
\label{Cosmological constraints and df}
As it was mentioned, we have  introduced an ansatz as equation \ref{eq:wDE_z_Ansatz}, that  consist of three free parameters. Where $\omega_1 $ indicates present time value of $\omega_{DE}$. For more convenience we can suppose $\omega_{\rm{0}}=f_0=1$ and therefore Eq.(\ref{eq:wDE_z_Ansatz}) is reduced to \cite{ch1:20a,Zim}
\begin{eqnarray} \label{eq:wDE_z_Ansatz2}
\omega_{DE}(z)=\omega_1 (1+z)^\beta .
\end{eqnarray}
Also, the mean square of relative error functions $\chi^2$, normally cause the free parameters plane split in  two parts. People usually are interested to the regions which  $\chi^2 / N \leq 1$, where $N$ denotes the amount of observational data. Whereas we use $Union-2$ data set for SNeIa, $N$ for supernovae is   $N_{SNe}=557,$ and also for OHD, CMB and BAO, one has $N_{OHD}=28$,  $N_{CMB}=1$ and  $N_{BAO}=1$. Since in this work three free parameters are appeared, the space of constraints has three dimensions. Thence for better expression, one can map figures on two dimensions (in fact it is supposed that, the free parameters are independent) and their values will be analyzed. The common regions for best fitting of all constraints,  play key role in this study. Based on above discussions we plot a couple of free parameters in Figures \ref{fig:OM1} - \ref{fig:B1}.  In Figure \ref{fig:OM1} we  investigate the constraints on $\Omega_{m0}$ in $\omega_1 \, \beta$ plate, and also for two constraints SNeIa and OHD minimum points of $\chi^2 $ are distinguished. In Figures \ref{fig:w1}   using best value of $\omega_1$, the constraints in $\Omega_{m0} \, \beta$ are obtained, in a similar way for best value of $\beta$, the behavior  of constraints in $\omega_1 \, \Omega_{m0} $ surface will be shown.  Let us, return our attention to figure \ref{fig:OM1} again. For $\Omega_{m0}=0.2$, the CMB, BAO and OHD have an overlap region, but they are not in agreement with SNeIa results. Also for a different quantity, the SNeIa and OHD results could be in agreement with together. This different  behavior of constraints indicates that if one wants to compare theoretical results with observations, it is better the greatest set of constraints, to be considered. For more investigation about overlaps and the effects of individual observations, we plot figures \ref{Projection2DSNeCMBBAO} and \ref{Projection2DCMBBAOOHD}. In figure \ref{Projection2DSNeCMBBAO} the behaviour of  $\chi_{\rm T} ^2 = \chi_{\rm{SNe}} ^2 + \chi_{\rm{OHD}} ^2 + \chi_{\rm{CMB}} ^2 + \chi_{\rm{BAO}} ^2$ and $\chi_{\rm T} ^2 = \chi_{\rm{SNe}} ^2 + \chi_{\rm{CMB}} ^2 + \chi_{\rm{BAO}} ^2$ for $\Delta \chi_{\rm T} ^2 = 3.53, 6.25, 8.02$ are compared. Also in figure \ref{Projection2DCMBBAOOHD} to investigate degeneracy one can consider  $\chi_{\rm T} ^2 = \chi_{\rm{SNe}} ^2 + \chi_{\rm{OHD}} ^2 + \chi_{\rm{CMB}} ^2 + \chi_{\rm{BAO}} ^2$ and $\chi_{\rm T} ^2 = \chi_{\rm{OHD}} ^2 + \chi_{\rm{CMB}} ^2 + \chi_{\rm{BAO}} ^2$ for $\Delta \chi_{\rm T} ^2 = 0.1, 0.2, 0.3$. These two figures indicate that although the importance of individual OHD data surveying (in comparison SNeIa, CMB and BAO) is not so important, but it decreases the degeneracy between free parameters of the model. From figures \ref{Projection2DSNeCMBBAO} and \ref{Projection2DCMBBAOOHD}, it is obvious that a collective of four constraints has completely different results in comparison to even three constraints. In the following,  by means of observations, we use some custom quantities which are considered for better estimation of theoretical parameters of the model. Since all free parameters of the model are independent, the total likelihood function could be introduced as
\begin{eqnarray}\label{total likelihood L_T}
{\cal L}_{\rm{T}}  = {\cal L}_{{\rm{SNe}}}  \times {\cal L}_{{\rm{OHD}}}  \times {\cal L}_{{\rm{CMB}}}  \times {\cal L}_{{\rm{BAO}}},
\end{eqnarray}
therefore the total $\chi^2$ function  could be achieved as
\begin{eqnarray}\label{total chi2}
\chi _{\rm{T}}^2  = \chi _{{\rm{SNe}}}^2  + \chi _{{\rm{OHD}}}^2  + \chi _{{\rm{CMB}}}^2  + \chi _{{\rm{BAO}}}^2.
\end{eqnarray}
It is considerable  to attain the maximum amount of the probability and  the minimum value of $\chi^2$, we should minimize $\chi _{\rm{T}}^2$. Also it should be noted, in \ref{total chi2} all components have same weight. So the likelihood method is equivalent to this fact that, for instance all measurements which lead to CMB is equal to a supernova explosion!. Afterwards we return to this problem. Another quantity which could be used for data fitting process is
\begin{equation}\label{tilde_{chi2}}
\tilde{\chi}^{2}=\frac{\chi_{T}^2}{N_{dof}}
\end{equation}
where subscript $dof$ is abbreviation of degree of freedom, and $N_{dof}$ could be defined as the difference between all observational sources and the amount of free parameters. Let's explain it in more detail, whereas the amounts of all observations are $557+28+1+1 = 587$, and the number of free parameters are $4$, by considering $H_0$, therefore $N_{dof}$, is equal to $583$.
Also one knows, the acceptable quantity for $\tilde \chi ^2 $ is $1.05$. For more convenient, we now define the average relative error functions as follows
\begin{eqnarray}\label{tilde{chi^2} for any obs}
 &\bar \chi _{{\rm{SNe}}}^2  = \frac{{\chi _{{\rm{SNe}}}^2 }}{{N_{{\rm{SNe}}} }}, \\
& \bar \chi _{{\rm{OHD}}}^2  = \frac{{\chi _{{\rm{OHD}}}^2 }}{{N_{{\rm{OHD}}} }}, \\
 &\bar \chi _{{\rm{CMB}}}^2  = \frac{{\chi _{{\rm{CMB}}}^2 }}{{N_{{\rm{CMB}}} }}, \\
 &\bar \chi _{{\rm{BAO}}}^2  = \frac{{\chi _{{\rm{BAO}}}^2 }}{{N_{{\rm{BAO}}} }}.
\end{eqnarray}
Finally we can introduce $\chi_{\rm m} ^2$ function, which is equal to maximum of $\bar{\chi} ^2$ functions and it could be considered as
\begin{eqnarray}\label{chi_{m} ^2}
&\chi _{\rm{m}}^2  = {\rm{max}}\,{\rm{of}}\,\left( {\bar \chi _{{\rm{SNe}}}^2 ,\bar \chi _{{\rm{OHD}}}^2 ,\bar \chi _{{\rm{CMB}}}^2 ,\bar \chi _{{\rm{BAO}}}^2 } \right).
\end{eqnarray}
In fact the $\chi_{\rm m} ^2$ function could be considered as a criterion of accuracy for  the models. Now we want to compare the behavior of  $\chi_{\rm m} ^2$ and $\tilde{\chi} ^2$ functions. Without loss the generality of the model, one can plot the three dimensional shape of $\chi_{\rm m} ^2$ and $\tilde{\chi} ^2$, versus free parameters of the model. These diagrams help us to find out the best estimation of the free parameters in comparison to observations; for more clarity one can see the figure \ref{fig:Projection To D1}. In this figure, the first diagram shows  the minimum of $\chi_{\rm m} ^2$ and $\bar{\chi} ^2$ versus $\Omega _{m0}$. Also in two latest diagrams of figure \ref{fig:Projection To D1}, the minimum points are drown based on $\omega_1$ and $\beta$ respectively. By comparison  the behavior of these relative error functions in Figure \ref{fig:Projection To D1} one can realize that, there are more points (or neighborhood) in which  $\tilde{\chi} ^2 < 1$, but $\chi_m ^2$ exceeds $1.05$. In fact this behavior was predictable, because in definition of $\tilde{\chi} ^2$, we use the contribution of all observational data set. So, for example the $\chi _{{\rm{CMB}}}^2$ deviation of best fitting results, could be recompense by SNeIa data abundance. We will return to this drawback, after some discussion about likelihood and relative error functions.
For more illustration, we portrait the different surfaces of three dimensional, $(\Omega_{\rm{m0}}, \omega_{\rm{1}}, \beta)$, to $(\Omega_{\rm{m0}}, \beta)$, $(\omega_1 ,\Omega_{\rm{m0}})$ and $(\omega_1 ,\beta)$ surfaces, which are brought in Figures \ref{Fig:Projection2D_w}, \ref{Fig:Projection2D_b} and \ref{Fig:Projection2D_OM}. Diagrams  $B$ and $C$ are related to  $(\chi_{\rm T} ^2)_{\rm{min}}$, where the subscript $min$, shows the minimum value of $\chi _{{\rm{T}}}^2$ . It is notable, in a three dimensional space of free parameters, the confidence levels $68.3\%$, $90\%$ and $95.4\%$  are proportional to $\Delta \chi_{\rm T} ^2 = 3.53$,  $\Delta \chi_{\rm T} ^2 = 6.25$ and $\Delta \chi_{\rm T} ^2 = 8.02$ surfaces respectively where $\Delta \chi_{\rm T} ^2  =\chi_{\rm T} ^2  - (\chi_{\rm T} ^2)_{\rm{min}}$. In diagram $B$ of  Figures \ref{Fig:Projection2D_w}, \ref{Fig:Projection2D_b} and \ref{Fig:Projection2D_OM} the  counter lines of confidence levels are drown and in diagram $C$, both the $\chi_{\rm m} ^2$ surfaces and counter lines are brought for more comparison. From diagram $C$ it is realized that the confidence level counters exceed the $\chi_{\rm m} ^2$ regions. From this behaviour it is concluded that, the theoretical prediction of CMB shift parameter is very greater than it's observational quantity. As mentioned heretofore, when  the total mean square error function is introduced the weight of all constraints was identical and this causes some problems.
As a matter of fact, the results of likelihood's parameter, equation \ref{total chi2},   the effect of CMB shift parameter in comparison to the abundant  SNeIa data set is ignored. For more information,  one can see Table \ref{tab:Chi_T_min} and definition of $N_{dof}$. To overcome these problems, we redefine $\chi _{\rm{T}}^2$ as bellow
 \begin{eqnarray}\label{chi_T2 re-weight}
\chi _{\rm{T}}^2  = \chi _{{\rm{SNe}}}^2  + \chi _{{\rm{OHD}}}^2  + 3 \, \chi _{{\rm{CMB}}}^2  + 3 \, \chi _{{\rm{BAO}}}^2.
\end{eqnarray}
It should be noted, in data fitting and maximization of probability quantities these two definitions of $\chi_{\rm T} ^2$, i.e. equations~\ref{total chi2} and \ref{chi_T2 re-weight}, have not very different. For justifying this claim one can compare Tables \ref{tab:Chi_T_min} and \ref{tab:Chi_T_min_w} which are related to \ref{total chi2} and \ref{chi_T2 re-weight} respectively. But in  figures which related to confidence levels one can observe that the exceeding of confidence levels are reduced, therefore the re-weight of some constraints can improve the behaviour of the model.  For more clarification one can refer to figures \ref{Fig:WeProjection2D_w},  \ref{Fig:WeProjection2D_b} and \ref{Fig:WeProjection2D_OM}. Now by means of   \ref{chi_T2 re-weight}, we margin the likelihood $\mathcal{L} (\Omega_{\rm{m0}} ,\omega_1 , \beta)$ w.r.t $\omega_1$, $\beta$ and $\Omega_{\rm{m0}}$ respectively. Also the relative probability functions  $\mathcal{L} (\Omega_{\rm{m0}}, \beta)$, $\mathcal{L} (\omega_1, \Omega_{\rm{m0}})$ and  $\mathcal{L} (\omega_{1} , \beta)$ in  two dimensional confidence levels $68.3\%$, $90\%$ and $95.4\%$ are plotted  in Figure \ref{Fig:We_3CMB_3BAO_Marginalized_2D}. For more investigations, we will draw the one dimensional marginalized likelihood functions $\mathcal{L}(\Omega_{\rm{m0}})$ versus $\Omega_{\rm{m0}}$, $\mathcal{L}(\omega_1)$ based on $\omega_1$ and $\mathcal{L}(\beta)$ versus $\beta$ in figure \ref{Fig:We_MarginalizedTo1D}. Meanwhile in Table \ref{tab:Moste_P_marginalized} one observes the quantities which maximize the marginalized likelihoods using different confidence levels by means of confidence levels $\sigma_1 = 68.3\%$ and $\sigma_2 = 95.4\%$ .
\subsection{Typical example}
\label{Typicalexample}
\begin{figure}
\includegraphics [width=6cm]{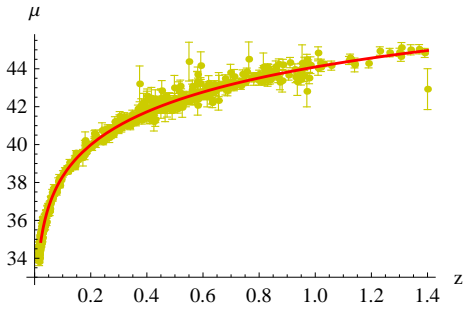}
\caption{\small{The observed distance modulus of supernovae (points) and the theoretical predicted distance
modulus (red-solid line) in the context of ADE model.}}
\label{Fig5}
\end{figure}

 Now,  we define an effective dark energy as combination of dark energy $\rho_{de}$ and scalar field density as $\rho_{DE}=\rho_{de}+{\rho_{\varphi}}/{f(\varphi)}$. So, the Friedmann equation is rewritten as
 \begin{equation}\label{14}
3H^2=f(\varphi) \Big( \rho_m + \rho_{DE} \Big).
\end{equation}
An useful parameter in this study is energy density parameter $\Omega$. Here $\Omega_{DE}$ and $\Omega_m$ respectively will be taken equal to $\Omega_{DE}=f(\varphi)\rho_{DE}/ \rho_c $ and $\Omega_m=f(\varphi) \rho_m/ \rho_c$, in which $\rho_c$ is the critical energy density which is defined as $\rho_c=3H^2$. As a result, from the Friedmann equation we have $\Omega_{DE}+\Omega_m=1$.\\
To obtain energy conservation equations for effective dark energy one can achieve the following results
\begin{equation}\label{15}
\frac{d}{dt}{\Big(f(\varphi)\rho_{DE}\Big)}+3Hf(\varphi)(1+\omega_{DE})\rho_{DE} = \gamma \rho_m\dot{f}(\varphi),
\end{equation}
\begin{equation}\label{16}
\frac{d}{dt}{\Big(f(\varphi)\rho_m\Big)}+3Hf(\varphi)(1+\gamma)\rho_m = -\gamma \rho_m\dot{f}(\varphi),
\end{equation}
so that the effective pressure of dark energy is defined as $p_{DE}=p_{\Lambda}+{p_{\varphi}}/{f(\varphi)}$, and one has the effective dark energy equation of state parameter as $\omega_{DE}={p_{DE}}/{\rho_{DE}}$. Also $\gamma$ is the matter equation of state parameter which is defined as $\gamma={p_m}/{\rho_m}$. For $\gamma=constant$,   integrating of Eq.(\ref{16}) results in the following relation for cold dark matter energy density
\begin{equation}\label{17}
\rho_m=\frac{\rho_{em}^0}{a^{3(1+\gamma)}f^{(1+\gamma)}(\varphi)},
\end{equation}
where $\rho_{em}^0= f_{0}^{(1+\gamma)}(\varphi)\rho_{m}^0$.
In this step, we suppose that the effective dark energy could be defined as ADE, in other word we assume that
\begin{equation}\label{18}
\rho_{DE}\equiv \rho_{ADE}=\frac{3n^2}{T^2},
\end{equation}
where $n$ is a numerical constant and $T$ is cosmic time and therefore $\Omega_{DE}$ is obtained as $\Omega_{DE}={f(\varphi)n^2}/{H^2T^2}$.  Taking this assumption and using  Eq.(\ref{15}), the equation of state parameter of effective dark energy could be acquired as
{\begin{equation}\label{19}
\omega_{DE}=-1 + \frac{2}{3}\frac{1}{n}\sqrt{\frac{\Omega_{DE}}{f(\varphi)}} + \frac{\dot{f}(\varphi)}{3Hf(\varphi)} \Big( \gamma r - 1 \Big),
\end{equation}}
where $r$ is ratio of cold dark matter and effective dark energy, namely $r={\rho_m}/{\rho_{DE}}={\Omega_m}/{\Omega_{DE}}$. The interaction term in this model generates an extra term for $\omega_{DE}$, which can justify the phantom divide line crossing.
By definition  an ansatz for $\omega_{e \Lambda}$, it can be considered as
\begin{equation}\label{19'}
\omega_{e \Lambda}+1=\omega_{0}+\omega_{1}(1+z)^{\beta}.
\end{equation}
For fitting the free parameters for ADE in an external scalar field interaction model, we use the $557$ Union-2 sample database of $SNeIa$, and   $\rho_m= \rho_{radiation}+\rho_{baryon}+\rho_{dark matter} $. Therefore in this case the Friemann equation is as
\begin{equation}\label{32}
3H^2=f(\varphi) \Big( \rho_m + \rho_{DE} \Big).
\end{equation}
Combining Eqs.(\ref{15})-(\ref{18}), give
\begin{equation}\label{33}
3H^2=f(\varphi) \Big(\frac{\rho^{0}_{em}}{a^{3(1+\gamma)}f^{(1+\gamma)}(\varphi)} + \frac{3n^2}{T^2} \Big),
\end{equation}
where $\rho^{0}_{em}$ is the effective energy density of matter at the present time. Whereas the
$557$ Union-2 sample database have collected from red shift parameter to various $SNeIa$, therefore we rewrite  $E=H/H_0$ versus $z$ as
\begin{equation}\label{34}
E^2=\frac{r_0(1+z)^3+(1+z)^{3\omega_0} \exp{\Big\{3\frac{\omega_1}{\beta}}[(1+z)^{\beta}-1]\Big\}}{r_0+1}.
\end{equation}
 To achieve the best fit for free parameters based on subsection {\ref{Supernovae type Ia}} a minimization method leads to
\begin{eqnarray}\label{37'}
\chi^{2}_{sn_{min}}(\omega_0&=&1.1;~ \omega_{1}=-1.65; ~\beta=-2.25),\\
 \chi _{min}^2  &=& A - \frac{{B^2 }}{C}=542.75, \\
  \mu _0  &= & - \frac{B}{C}= 43.1089.
\end{eqnarray}
where implies ${\chi^{2}_{sn}}/{dof}$=
$\chi^{2}_{sn_{min}}/dof= 0.981 (dof = 553)$.
\\
In figure {\ref{Fig5}}, we show a comparison between theoretical
distance modulus and observed distance modulus of supernovae data. The red-solid
line indicates the theoretical value of distance modulus, $\mu_{th}$, for the best value of
free parameters $\omega_{1}=-1.65$, $\omega_{0}=1.1$ and $\beta=-2.25.$
 This shows that the
model is clearly consistent with the data since $\chi^{2}/dof = 1$.
Fig.2 show contour plots for the free parameters $\omega_{1}$ and $\beta$,  it is shown that the best value for these parameters are $-1.86<\omega_{1}<-1.62$ and  $-2.27<\beta<-0.73$ in which for  stability condition  $c^2 >0$, we have taken the interface between green and yellow sector, $\omega_1=-1.68$.\\
\begin{figure}\label{Fig6}
\includegraphics[width=6cm]{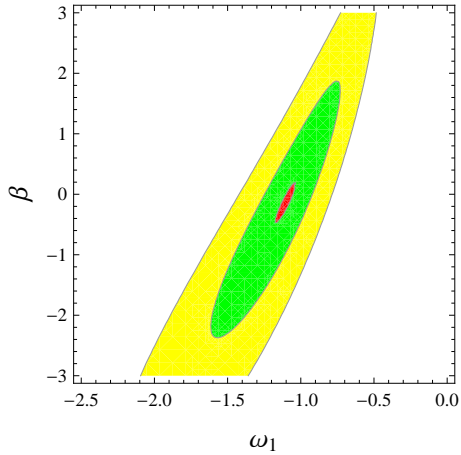}
\caption{ {\small Contour plots for the free parameters $\omega_{1}$ and $\beta$, shows that the best value for these parameters are $-1.86<\omega_{1}<-1.62$ and  $-2.27<\beta<-0.73$.}}
\end{figure}\\
The evolution of effective dark energy parameter, $\omega_{DE}$, versus $z$, for $\omega_0= 1.1$, $\omega_1=-1.68$ and $\beta =-2.25 $ have been shown in  Fig.3. This show that by growing  $z$  the parameter get into the phantom phase.
\begin{figure}[h]\label{Fig1'}
\includegraphics[width=7cm]{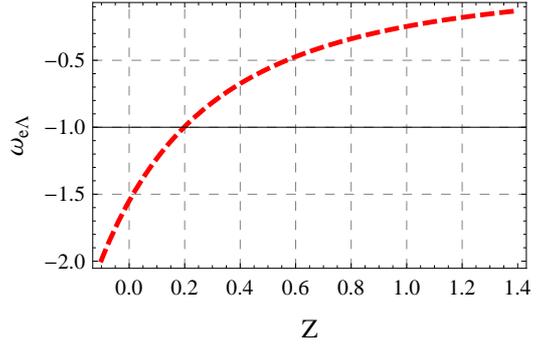}
\caption{ {\small The plot shows the evolution of effective dark energy parameter, $\omega_{DE}$, versus $z$, for $\omega_0=1.1 $, $\omega_1=-1.68$ and $\beta =-2.25 $. }}
\end{figure}\\
Here $\omega_{0}$, $\omega_{1}$ and $\beta$ are free parameters of the model which obtained from data fitting. It is clear that, if the form of dark energy density is given the coupling function, $f(\varphi)$, could be easily determined. For instance by using  Eqs.(\ref{15}), (\ref{16}) and (\ref{18}) one can obtain
\begin{equation}\label{19a}
f(\varphi)=f_{0} t^{2} a^{-3\omega_{0}} \exp{\Big[3\omega_{1}\frac{{(z+1)}^{\beta+2}}{\beta+2}\Big]},
\end{equation}
here $f_{0}$ is the constant of integration.
Whereas $$\frac{\dot{f}(\varphi)}{f(\varphi)}=3H\big[\frac{2}{3tH} -\omega_{0}-\omega_1{(1+z)}^{\beta+2}\big].$$
A significant  result of observational data is accelerated expansion of the Universe. A good cosmological model should be able to describe this acceleration. An useful quantity to investigate this property of the Universe is deceleration parameter which defined as $q=-1-{\dot{H}}/{H^2}$.
Using Eqs.(\ref{14}), (\ref{15}) and (\ref{16}), one achieves the deceleration parameter gives
\begin{eqnarray}\label{23}\nonumber
q&= -1+\frac{3}{2}\Big[1-\omega_0-\omega_{1}{(1+z)}^{\beta}\Big]\times\\ &
\Big(\frac{D_{0}}{1+(1+z)^{3(1-\omega_{0})}(t)\exp[\frac{-3\omega_1 {(1+z)}^{\beta}}{\beta}]}\Big),
\end{eqnarray}
where $D_{0}$ is the constant of integration. It is clearly seen that for  $\omega_0=1.1,\quad \omega_1=-1.68,\quad \beta=-2.25$, (which have obtained from data fitting processes)   $q<0$.

\section{Conclusion and Discussion}
\label{Conclusion}
\begin{figure}
\includegraphics[width=5cm]{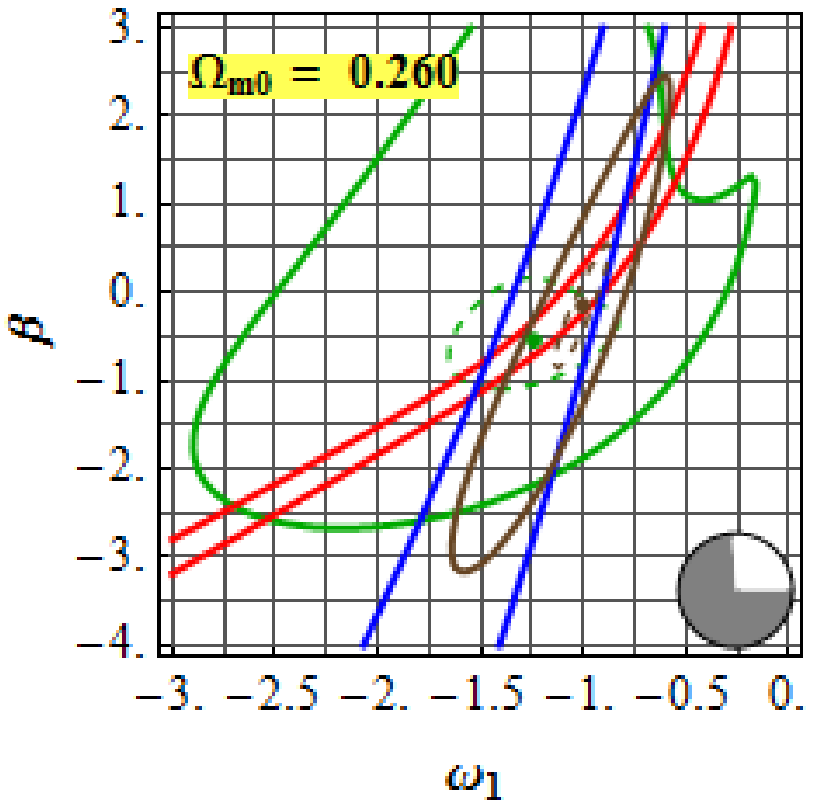}
\caption{\small{The counter lines of  $\chi^2 _{SNe} = 557$ (brown), $\chi^2 _{OHD} = 28.0$ (green), $\chi^2 _{CMB} = 1.0$ (red),  and  $\chi^2 _{BAO} = 1.0(blue)$ for  of $\Omega _{m0}=0.26$ are plotted. Also for two constraints $\rm{SNeIa}$ and $\rm{OHD}$  minimum points of $\chi^2$   are distinguished. The dashed lines refer to the counter lines which are greater of the minimum points only unity. }}
\label{fig:OM1}
\end{figure}

\begin{figure} %
\includegraphics[width=5cm]{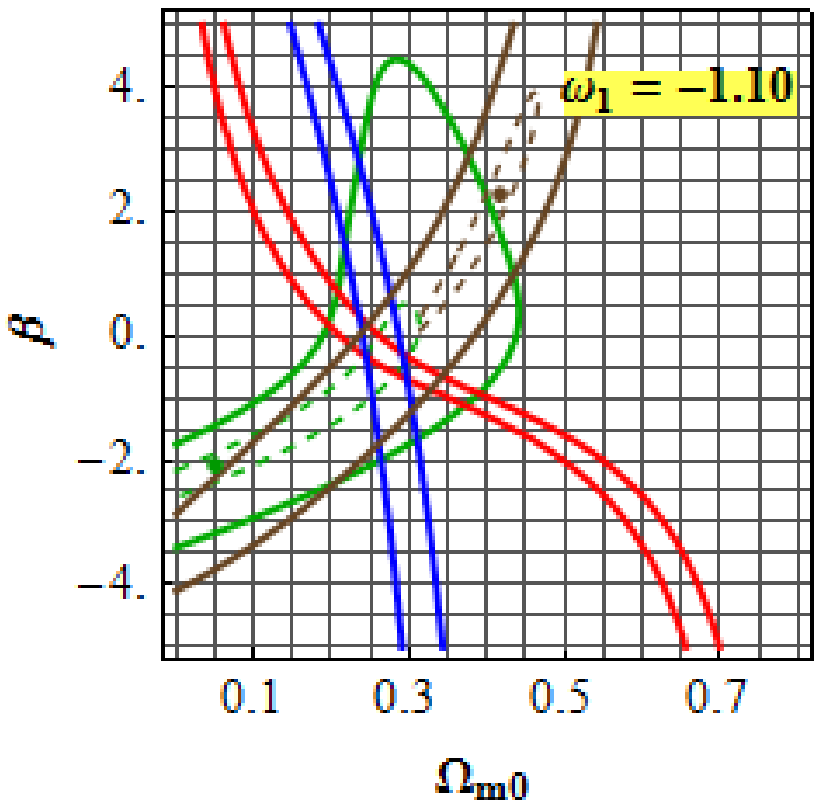}
\caption{\small{The counter lines of  $\chi^2 _{SNe} = 557$ (brown), $\chi^2 _{OHD} = 28.0$ (green), $\chi^2 _{CMB} = 1.0$ (red),  and  $\chi^2 _{BAO} = 1.0(blue)$ for $\omega _{1}=-1.1$ are plotted. Also for two constraints $\rm{SNeIa}$ and $\rm{OHD}$  minimum point of $\chi^2$  are distinguished. The dashed lines refer to the counter lines which are greater of the minimum points only unity.}}
\label{fig:w1}
\end{figure}

\begin{figure}
\includegraphics[width=5cm]{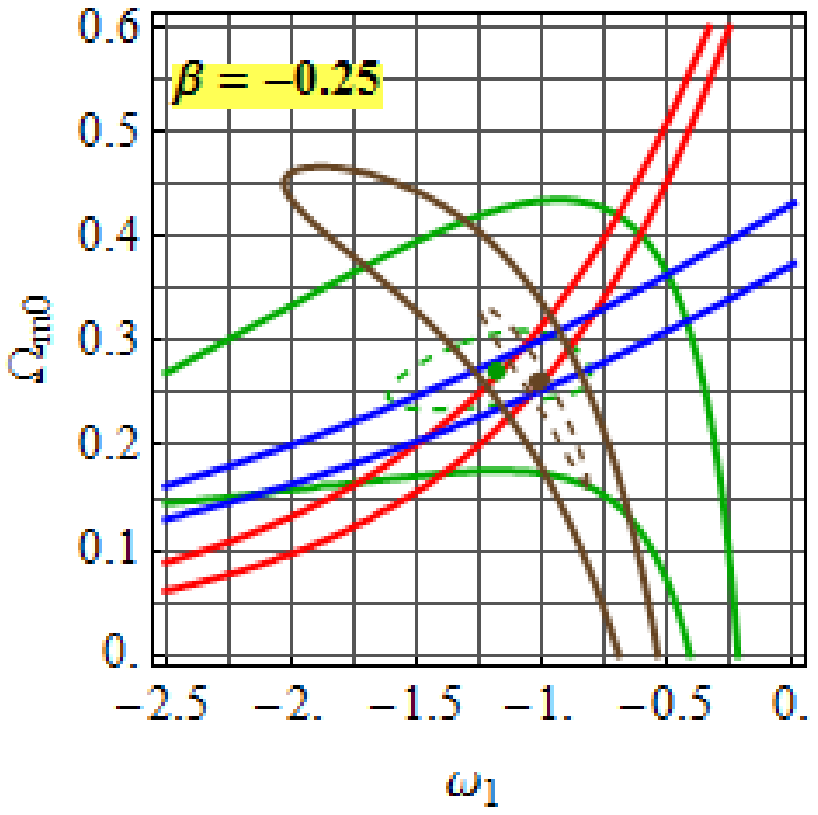}
\caption{\small{The counter lines of  $\chi^2 _{SNe} = 557$ (brown), $\chi^2 _{OHD} = 28.0$ (green), $\chi^2 _{CMB} = 1.0$ (red),  and  $\chi^2 _{BAO} = 1.0(blue)$ for $\beta=-0.25$ are plotted. Also for two constraints $\rm{SNeIa}$ and $\rm{OHD}$  minimum point of $\chi^2$   are distinguished. The dashed lines refer to the counter lines which are greater of the minimum points only unity.}}
\label{fig:B1}
\end{figure}

\begin{figure}
\includegraphics[width=7cm]{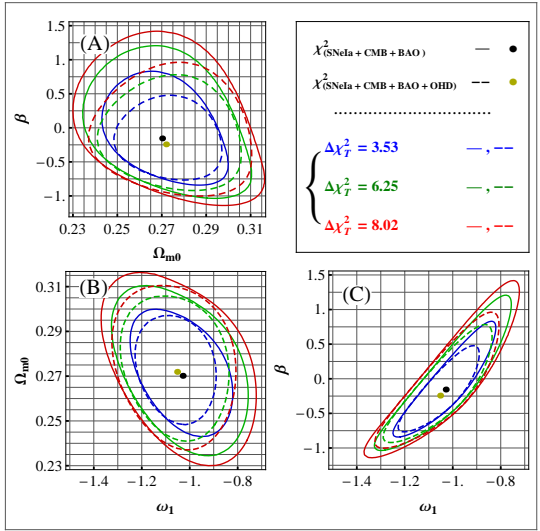}
\caption{\small{In this figure the behaviour of  $\chi_{\rm T} ^2 = \chi_{\rm{SNe}} ^2 + \chi_{\rm{OHD}} ^2 + \chi_{\rm{CMB}} ^2 + \chi_{\rm{BAO}} ^2$ (dashed counters)and $\chi_{\rm T} ^2 = \chi_{\rm{SNe}} ^2 + \chi_{\rm{CMB}} ^2 + \chi_{\rm{BAO}} ^2$ (solid counters) for $\Delta \chi_{\rm T} ^2 = 3.53 (inner\,loops), 6.25 (middle\,loops), 8.02 (outer\,loop)$ are compared. The minimum points of these two $\chi_{\rm T} ^2$ functions are distinguished by Solid points.}}
\label{Projection2DSNeCMBBAO}
\end{figure}

\begin{figure}
\includegraphics[width=7cm]{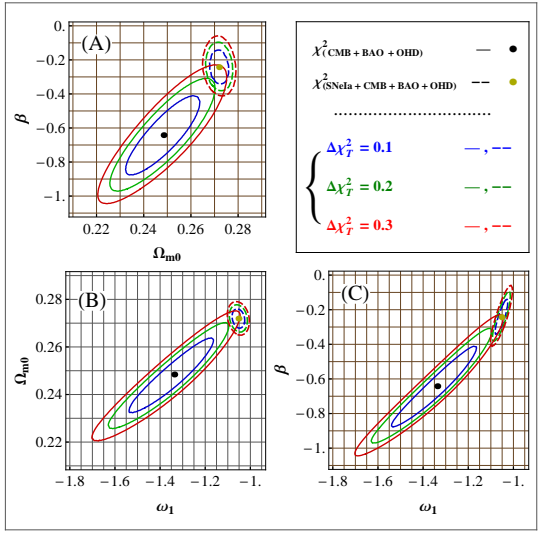}
\caption{\small{In this plot we consider  $\chi_{\rm T} ^2 = \chi_{\rm{SNe}} ^2 + \chi_{\rm{OHD}} ^2 + \chi_{\rm{CMB}} ^2 + \chi_{\rm{BAO}} ^2$ (dashed counters) and $\chi_{\rm T} ^2 = \chi_{\rm{OHD}} ^2 + \chi_{\rm{CMB}} ^2 + \chi_{\rm{BAO}} ^2$ (solid counters) for $\Delta \chi_{\rm T} ^2 = 0.1 (inner\,loops), 0.2 (middle\,loops), 0.3 (outer\,loop)$ to investigate degeneracy in this work. This Figure and Figure \ref{Projection2DSNeCMBBAO} indicate that although the importance of individual OHD data surveying in cosmological investigations (in comparison SNe Ia, CMB and BAO) is not so important but it causes decreasing degeneracy between free parameters of the model. The minimum points of these two $\chi_{\rm T} ^2$ functions are distinguished by Solid points.}}
\label{Projection2DCMBBAOOHD}
\end{figure}

\begin{figure} 
\includegraphics[width=7cm]{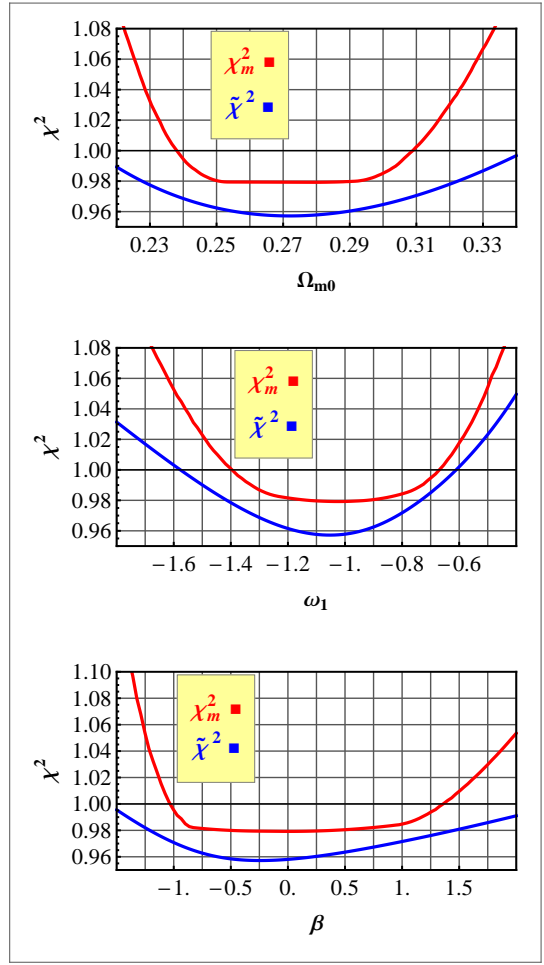}
\caption{\small{ In above diagrams the minimum quantity of $\tilde{\chi} ^2$ (blue line) and $\chi_m ^2$ (red-upper-line) versus
$\Omega_{\rm{m0}}$, $\omega_1$ and $\beta$ parameters have been drown respectively. }}
\label{fig:Projection To D1}
\end{figure}

\begin{figure} 
\includegraphics[width=7cm]{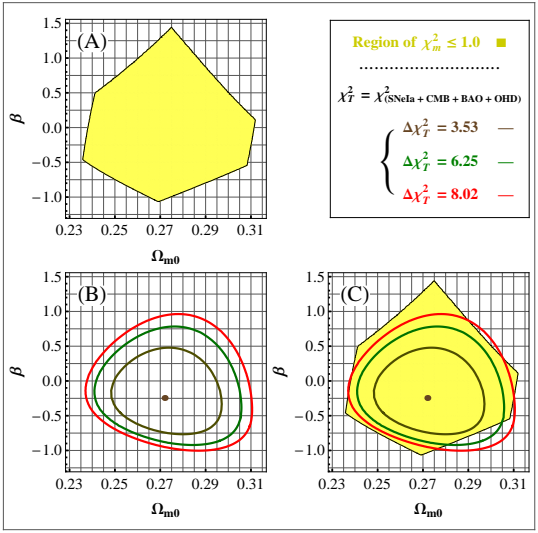}
\caption{\small{In diagram $(\rm A)$, the image of $\chi_{m} ^2 \leq 1$  on the $(\Omega_{m0}, \beta)$ surface are portrait. In $(\rm B)$ the minimum point of  $\chi_{\rm T} ^2 = \chi_{\rm{SNe}} ^2 +  \chi_{\rm{OHD}} ^2 +  \chi_{\rm{CMB}} ^2 +  \chi_{\rm{BAO}} ^2$ and the shadow of $\Delta \chi_{\rm T} ^2 = 3.53 (inner \,loop), 6.25 (middle\,loop), 8.02 (outer \,loop)$ surfaces on the $(\Omega_{m0}, \beta)$ plate,  are plotted. In part $(\rm C)$ both diagrams $(\rm A)$ and $(\rm B)$ are brought to compare the results.}}
\label{Fig:Projection2D_w}
\end{figure}

\begin{figure} 
\includegraphics[width=7cm]{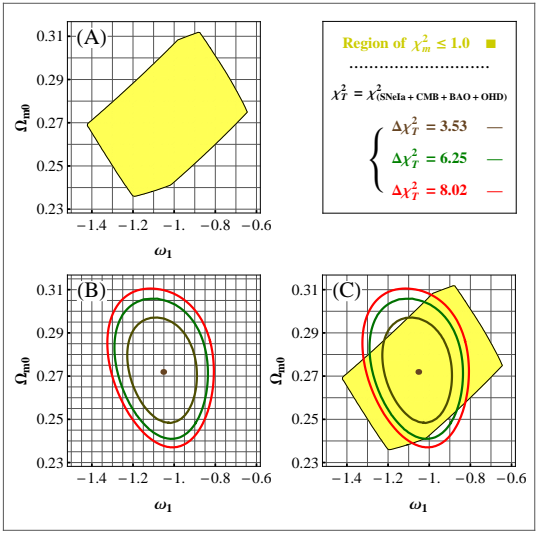}
\caption{\small{In diagram $(\rm A)$, the image of $\chi_{m} ^2 \leq 1$  on the $(\omega_1,\Omega_{m0})$ surface are portrait. In $(\rm B)$ the minimum point of  $\chi_{\rm T} ^2 = \chi_{\rm{SNe}} ^2 +  \chi_{\rm{OHD}} ^2 +  \chi_{\rm{CMB}} ^2 +  \chi_{\rm{BAO}} ^2$ and the shadow of $\Delta \chi_{\rm T} ^2 = 3.53 (inner\, loop), 6.25 (middle \,loop), 8.02 (outer \,loop)$ surfaces on the $(\omega_1,\Omega_{m0})$ plate, are drawn. In part $(\rm C)$ both diagrams $(\rm A)$ and $(\rm B)$ are considered for more comparison. }}
\label{Fig:Projection2D_b}
\end{figure}

\begin{figure} 
\includegraphics[width=7cm]{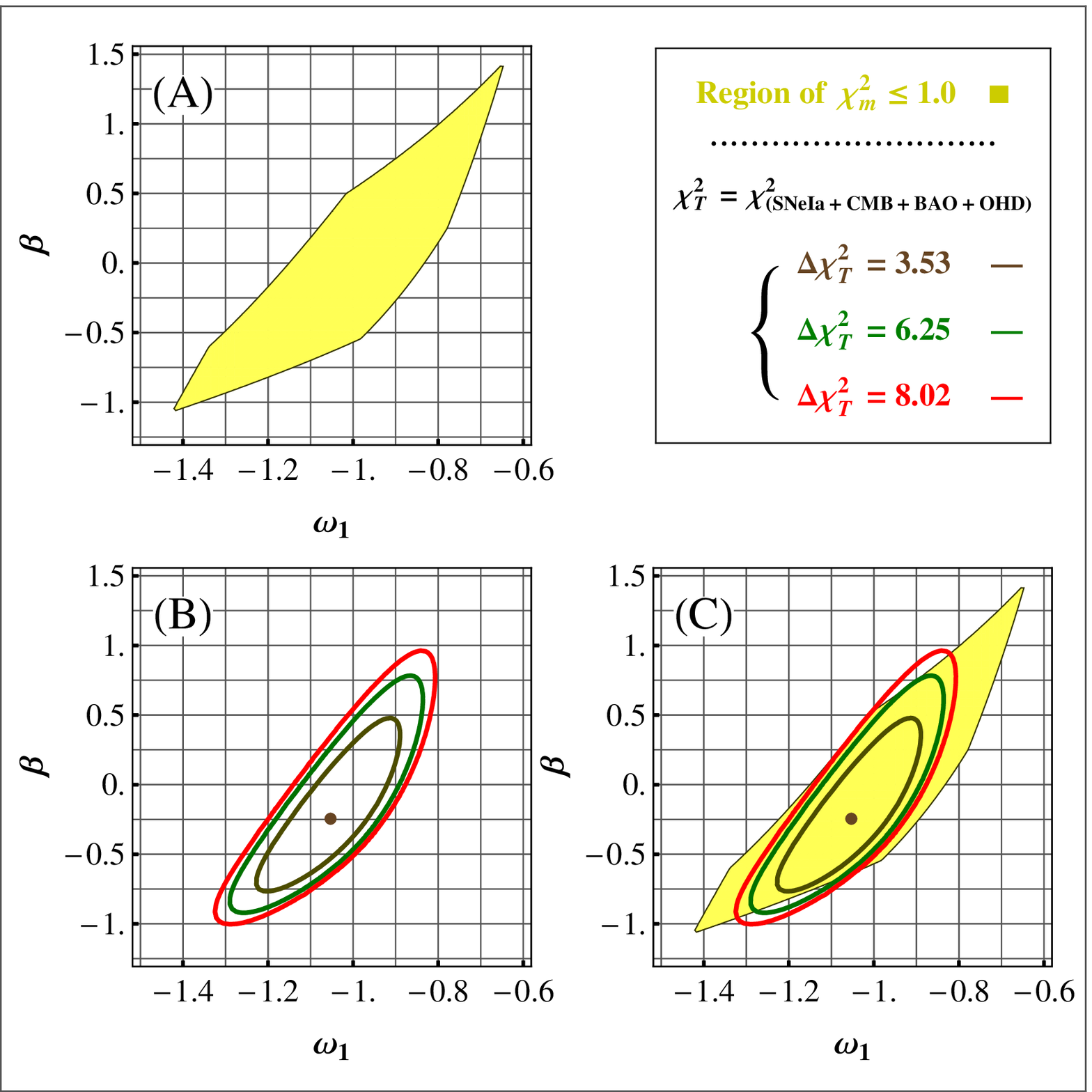}
\caption{\small{In diagram $(\rm A)$, the image of $\chi_{m} ^2 \leq 1$  on the $(\omega_1,\beta)$ surface are portrait. In $(\rm B)$ the minimum point of  $\chi_{\rm T} ^2 = \chi_{\rm{SNe}} ^2 +  \chi_{\rm{OHD}} ^2 +  \chi_{\rm{CMB}} ^2 +  \chi_{\rm{BAO}} ^2$ and the shadow of $\Delta \chi_{\rm T} ^2 = 3.53 (inner\, loop), 6.25 (middle\,loop), 8.02 (outer \,loop)$ surfaces on the $(\omega_1,\beta)$ plate , as counter lines, are drawn. In part $(\rm C)$ both diagrams $(\rm A)$ and $(\rm B)$ are compared.}}
\label{Fig:Projection2D_OM}
\end{figure}

\begin{figure} 
\includegraphics[width=7cm]{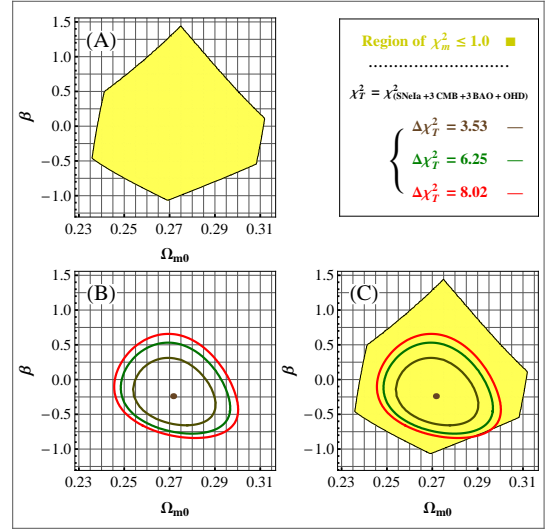}
\caption{\small{In diagram $(\rm A)$, the image of $\chi_{m} ^2 \leq 1$  on the $(\Omega_{m0}, \beta)$ surface are portrait. In $(\rm B)$ the minimum point of $\chi_{\rm T} ^2 = \chi_{\rm{SNe}} ^2 +  \chi_{\rm{OHD}} ^2 + 3 \chi_{\rm{CMB}} ^2 + 3 \chi_{\rm{BAO}} ^2$ and the shadow of $\Delta \chi_{\rm T} ^2 = 3.53 (inner\, loop), 6.25 (middle\, loop), 8.02 (outer\, loop)$ surfaces on the $(\Omega_{m0}, \beta)$ surface, as counter lines, are drawn. In part $(\rm C)$ both diagrams $(\rm A)$ and $(\rm B)$ have been brought for more comparison.}}
\label{Fig:WeProjection2D_w}
\end{figure}

\begin{figure} 
\includegraphics[width=7cm]{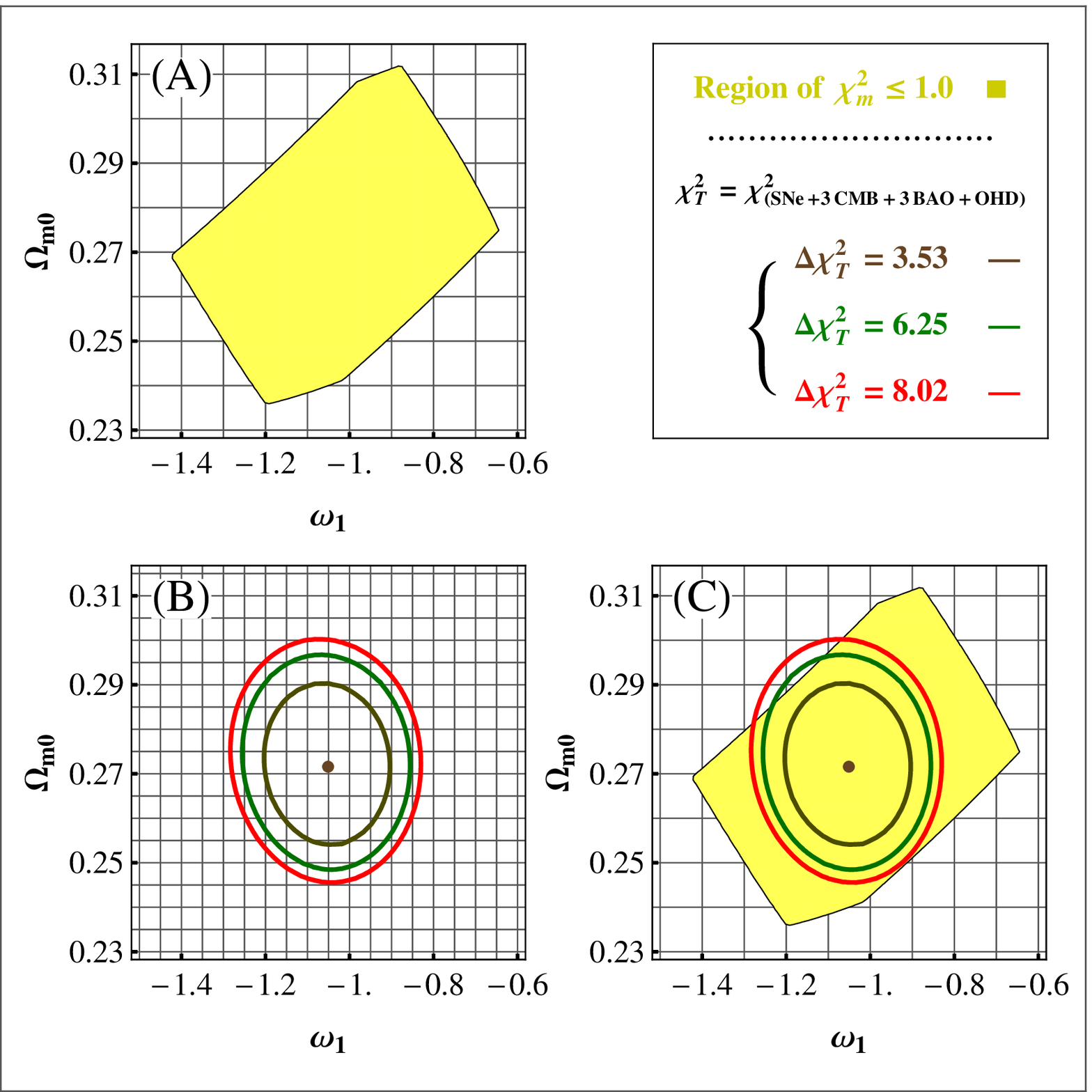}
\caption{\small{In diagram $(\rm A)$, the image of $\chi_{m} ^2 \leq 1$  on the $(\omega_1,\Omega_{m0})$ surface are portrait. In $(\rm B)$ the minimum point of $\chi_{\rm T} ^2 = \chi_{\rm{SNe}} ^2 +  \chi_{\rm{OHD}} ^2 + 3 \chi_{\rm{CMB}} ^2 + 3 \chi_{\rm{BAO}} ^2$ and the shadow of $\Delta \chi_{\rm T} ^2 = 3.53 (inner\, loop), 6.25 (middle\,loop), 8.02 (outer\, loop)$ surfaces on the $(\omega_1,\Omega_{m0})$ plate, as counter lines, are drawn. In part $(\rm C)$ both diagrams $(\rm A)$ and $(\rm B)$ are brought for comparison.}}
\label{Fig:WeProjection2D_b}
\end{figure}

\begin{figure} 
\includegraphics[width=7cm]{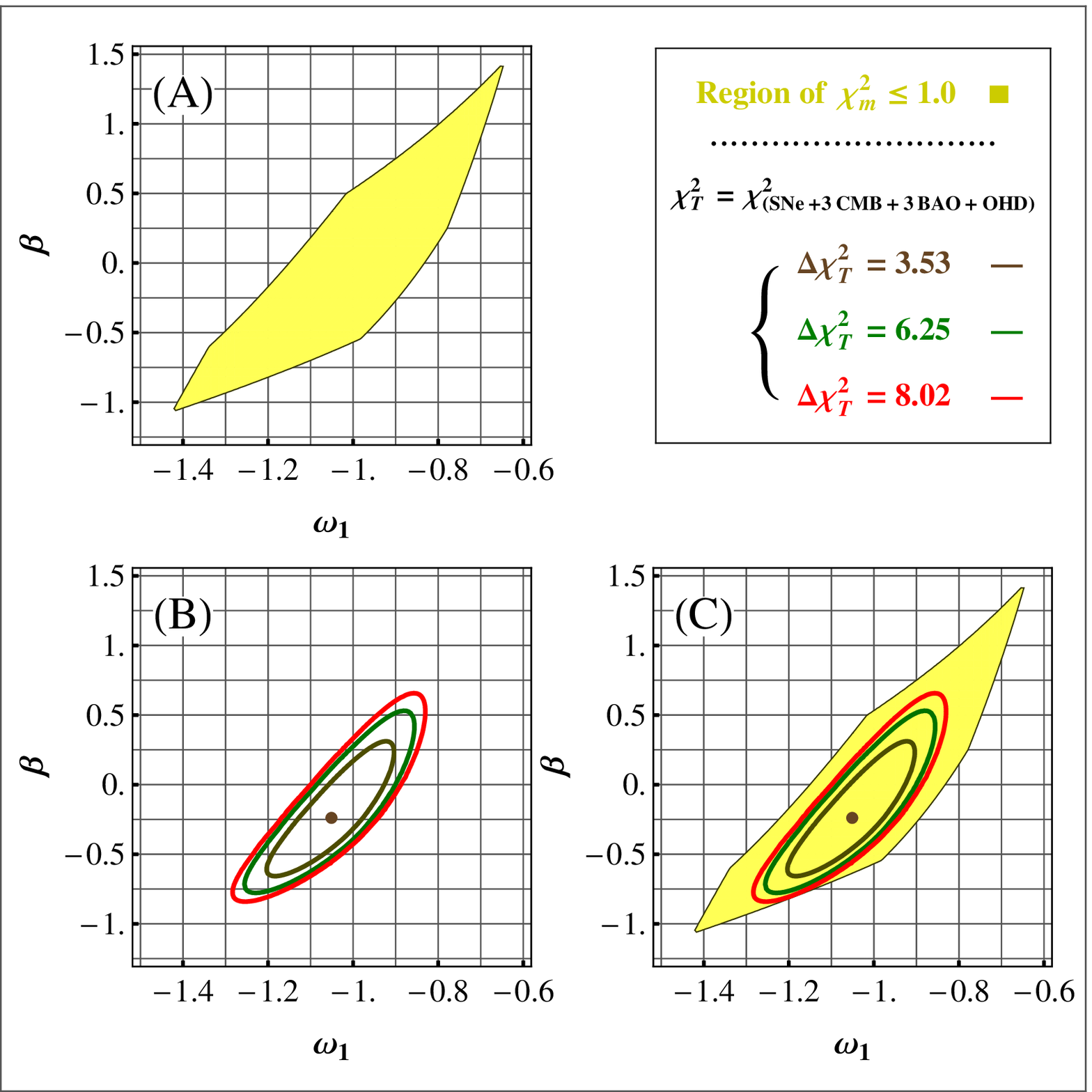}
\caption{\small{In diagram $(\rm A)$, the image of $\chi_{m} ^2 \leq 1$  on the $(\omega_1,\beta)$ surface are portrait. In $(\rm B)$ the minimum point of $\chi_{\rm T} ^2 = \chi_{\rm{SNe}} ^2 +  \chi_{\rm{OHD}} ^2 + 3 \chi_{\rm{CMB}} ^2 + 3 \chi_{\rm{BAO}} ^2$ and the shadow of $\Delta \chi_{\rm T} ^2 = 3.53 (inner\, loop), 6.25 (middle \,loop), 8.02 (outer\, loop)$ surfaces on the $(\omega_1,\beta)$ plate, as counter lines, are drawn. In part $(\rm C)$ both diagrams $(\rm A)$ and $(\rm B)$ are collected for comparison.}}
\label{Fig:WeProjection2D_OM}
\end{figure}

\begin{figure} 
\includegraphics[width=7cm]{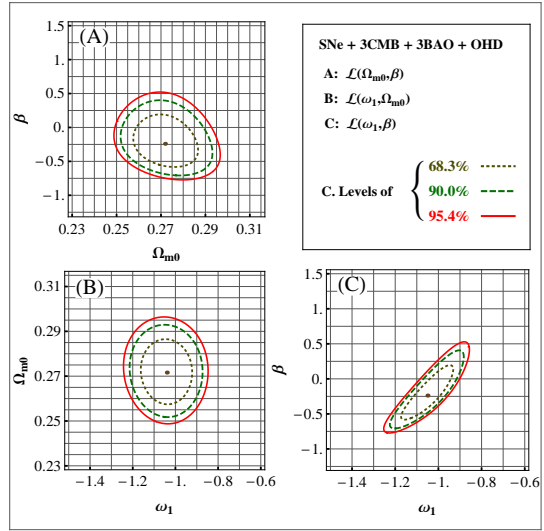}
\caption{\small{In the above two dimensional likelihood diagrams, the $68.3\%$ confidence level (dotted line), $90.0\%$ confidence level (green-dashed-line) and $95.45\%$ confidence level (red-solid-line) after marginalization on the  $\omega_1$, $\beta$ and  $\Omega_{m0}$ free parameters are plotted. Note that in this figure we use $\chi_{\rm T} ^2 = \chi_{\rm{SNe}} ^2 +  \chi_{\rm{OHD}} ^2 + 3 \chi_{\rm{CMB}} ^2 + 3 \chi_{\rm{BAO}} ^2$ and also the shadow of $\Delta \chi_{\rm T} ^2 = 3.53 (inner\, loop), 6.25 (middle \,loop), 8.02 (outer\, loop)$ surfaces.}}
\label{Fig:We_3CMB_3BAO_Marginalized_2D}
\end{figure}

\begin{figure} 
\includegraphics[width=7cm]{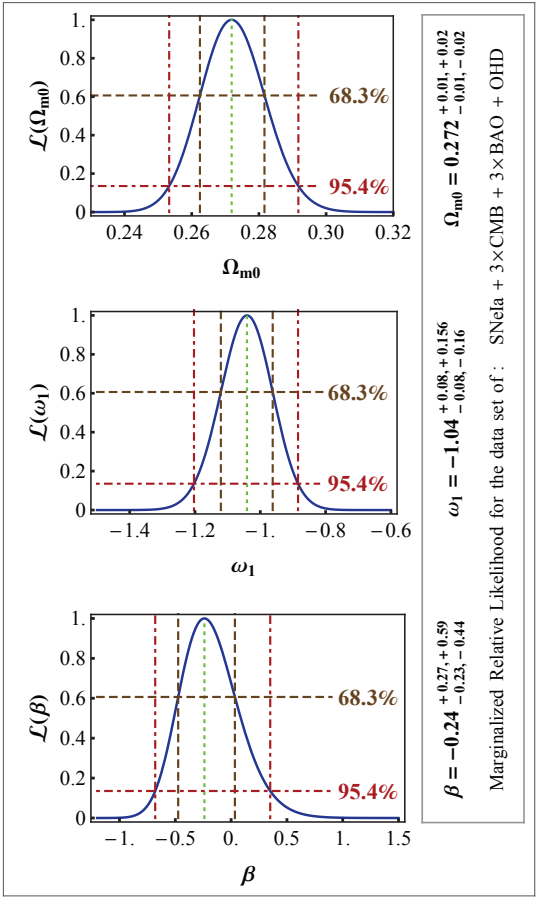}
\caption{\small{In above diagrams the relative likelihoods are plotted. The  $68.3\%$  and  $95.4\%$ confidence levels are distinguished as brown dashed line and red dot-dash line respectively.
It should be noted that in these figure we use $\chi_{\rm T} ^2 = \chi_{\rm{SNe}} ^2 +  \chi_{\rm{OHD}} ^2 + 3 \chi_{\rm{CMB}} ^2 + 3 \chi_{\rm{BAO}} ^2$. Also the best fits of the free parameters are as
$\beta = -0.24_{-0.23, - 0.44} ^ {+ 0.27, + 0.59}$,
$\omega_1 = -1.04_{-0.08, - 0.16} ^ {+ 0.08, + 0.156}$ and
$\Omega_{m0} =0.272_{-0.01, - 0.02} ^ {+ 0.01, + 0.02} $.}}
\label{Fig:We_MarginalizedTo1D}
\end{figure}

Interacting models which contain an external interaction between matter and scalar fields attract more attentions. Such mechanisms are capable to suppress the fifth force and also are in good agrement with observations. Using such powerful mechanism we have attained some cosmological parameters consist of coincidence and deceleration parameters. For instance based on Table {\ref{tab:Chi_T_min}}, and equations {\ref{eq:dr_dz}} and {\ref{eq:q_Ez}} it is clear that $r(z)$ is a decreasing function and $q$ has taken negative values for different amounts of $z$. Considering a suitable ansatz for EoS parameter of effective dark energy dimensionless Hubble parameter is obtained. So by means of SNeIa, CMB, BAO and OHD  data sets  the minimum value of $\chi^2$ for the free parameters of the model are achieved.
To estimate the free parameters of an ansatz for the effective dark energy equation of state, the whole of the observational data sets have been considered. For more details one can compare the results of figures \ref{Projection2DSNeCMBBAO} and \ref{Projection2DCMBBAOOHD}, and the results of typical example, subsection {\ref{Typicalexample}}. Also for getting better overlap between the counters with the constraint $\chi _{\rm{m}}^2\leq 1$, the  $\chi _{\rm{T}}^2$ function have been re-weighted. Meanwhile the relative probability functions have plotted for marginalized likelihood $\mathcal{L} (\Omega_{\rm{m0}} ,\omega_1 , \beta)$ according to two dimensional confidence levels $68.3\%$, $90\%$ and $95.4\%$. In addition the value of free parameters which maximize the marginalized likelihoods using above confidence levels have obtained. Based on above discussions a couple of free parameters in figures \ref{fig:OM1} - \ref{fig:B1}, have plotted. In figure \ref{fig:OM1}, the constraints on $\Omega_{m0}$ in $\omega_1 \, \beta$ plate have investigated; and also, for two constraints SNeIa and OHD minimum points of $\chi^2 $ have distinguished. In figures \ref{fig:w1},  using best value of $\omega_1$, the constraints in $\Omega_{m0} \, \beta$ surface are obtained. In a similar way, for best value of $\beta$, the behavior  of constraints in $\omega_1 \, \Omega_{m0}$ plane have shown. Also based on figure \ref{fig:OM1}, for $\Omega_{m0}=0.2$, the CMB, BAO and OHD have an overlap region, but they are not in agreement with SNeIa results; where one possible explanation would be an incompatibility among the data sets. Also, for a different values one can find a region which SNeIa and OHD are in more agreement against CMB and BAO. This different  behavior of constraints indicates that, if one wants to compare theoretical and observational results, it maybe better the greatest set of constraints to be considered. For more investigation about overlaps and the effects on individual observations, the figures \ref{Projection2DSNeCMBBAO} and \ref{Projection2DCMBBAOOHD}, have been plotted. In figure \ref{Projection2DSNeCMBBAO}, the behaviour of  $\chi_{\rm T} ^2 = \chi_{\rm{SNe}} ^2 + \chi_{\rm{OHD}} ^2 + \chi_{\rm{CMB}} ^2 + \chi_{\rm{BAO}} ^2$ and $\chi_{\rm T} ^2 = \chi_{\rm{SNe}} ^2 + \chi_{\rm{CMB}} ^2 + \chi_{\rm{BAO}} ^2$ for $\Delta \chi_{\rm T} ^2 = 3.53, 6.25, 8.02$, have compared. Also in figure \ref{Projection2DCMBBAOOHD}, we have considered $\chi_{\rm T} ^2 = \chi_{\rm{SNe}} ^2 + \chi_{\rm{OHD}} ^2 + \chi_{\rm{CMB}} ^2 + \chi_{\rm{BAO}} ^2$ and $\chi_{\rm T} ^2 = \chi_{\rm{OHD}} ^2 + \chi_{\rm{CMB}} ^2 + \chi_{\rm{BAO}} ^2$, for $\Delta \chi_{\rm T} ^2 = 0.1, 0.2, 0.3$, to investigate the degeneracy. These two figures indicate that although the importance of individual OHD data surveying in cosmological investigations (in comparison SNeIa, CMB and BAO) is not so important but it decreases degeneracy between free parameters.

\begin{acknowledgements}
H. Sheikhahmadi would like to thank Iran's National Elites Foundation for financially support during this work.
\end{acknowledgements}



\newpage

\begin{table}[tbp]
\caption{\small{In this table from left to right $z$, $H(z) (km\, s^{-1}\, Mpc ^{-1})$,  it's uncertainty $\sigma_{H} (km\, s^{-1}\, Mpc ^{-1})$ in measurement  and related references (by considering the technique which is used)  are collected, respectively.}}\label{tab: Hz}
\centering
\begin{tabular}{ccccc}
\hline
~~$z$&~~$H(z)$ &~~~$\sigma_{H}$&~References&~Techneque\\
\hline\hline
$0.070$&~$69$&~~$19.6$&~\cite{5}&~SDSS DR7; $0<z<0.4$\\
$0.100$&~$69$&~~$12$&~\cite{1}&~ATC; $0.1< z < 1.8$\\
$0.120$&~$68.6$&~~$26.2$&~\cite{5}&~SDSS DR7; $0<z<0.4$\\
$0.170$&~$83$&~~$8$&~\cite{1}&~ATC; $0.1< z < 1.8$\\
$0.179$&~$75$&~~$4$&~\cite{3}&~OHD+CMB; $0<z<1.75$\\
$0.199$&~$75$&~~$5$&~\cite{3}&~OHD+CMB; $0<z<1.75$\\
$0.200$&~$72.9$&~~$29.6$&~\cite{5}&~SDSS DR7; $0<z<0.4$\\
$0.270$&~$77$&~~$14$&~\cite{1}&~ATC; $0.1< z < 1.8$\\
$0.280$&~$88.8$&~~$36.6$&~\cite{5}&~SDSS DR7; $0<z<0.4$\\
$0.350$&~$76.3$&~~$5.6$&~\cite{7}&~SDSS DR7 LRGs; z=0.35\\
$0.352$&~$83$&~~$14$&~\cite{3}&~OHD+CMB; $0<z<1.75$\\
$0.400$&~$95$&~~$17$&~\cite{1}&~ATC; $0.1< z < 1.8$\\
$0.440$&~$82.6$&~~$7.8$&~\cite{6}&~WiggleZ+H(z); $z<1.0$\\
$0.480$&~$97$&~~$62$&~\cite{2}&~CMB+OHD; $0.2< z < 1.0$\\
$0.593$&~$104$&~~$13$&~\cite{3}&~OHD+CMB; $0<z<1.75$\\
$0.600$&~$87.9$&~~$6.1$&~\cite{6}&~WiggleZ+H(z); $z<1.0$\\
$0.680$&~$92$&~~$8$&~\cite{3}&~OHD+CMB; $0<z<1.75$\\
$0.730$&~$97.3$&~~$7.0$&~\cite{6}&~WiggleZ+H(z); $z<1.0$\\
$0.781$&~$105$&~~$12$&~\cite{3}&~OHD+CMB; $0<z<1.75$\\
$0.875$&~$125$&~~$17$&~\cite{3}&~OHD+CMB; $0<z<1.75$\\
$0.880$&~$90$&~~$40$&~\cite{2}&~CMB+OHD; $0.2<z< 1.0$\\
$0.900$&~$117$&~~$23$&~\cite{1}&~ATC; $0.1< z < 1.8$\\
$1.037$&~$154$&~~$20$&~\cite{3}&~OHD+CMB; $0<z<1.75$\\
$1.300$&~$168$&~~$17$&~\cite{1}&~ATC; $0.1< z < 1.8$\\
$1.430$&~$177$&~~$18$&~\cite{1}&~ATC; $0.1< z < 1.8$\\
$1.530$&~$140$&~~$14$&~\cite{1}&~ATC; $0.1< z < 1.8$\\
$1.750$&~$202$&~~$40$&~\cite{1}&~ATC; $0.1< z < 1.8$\\
$2.300$&~$224$&~~$8$&~\cite{4}&~BAO; $0.7<z<2.3$\\
\hline\hline
\end{tabular}
\end{table}


\renewcommand{\arraystretch}{1.1}
\begin{table}
\caption{\small{In the table, the quantities related to minimum point of
$\chi_{\rm T} ^2 = \chi_{\rm{SNe}} ^2 +  \chi_{\rm{OHD}} ^2 +  \chi_{\rm{CMB}} ^2 +  \chi_{\rm{BAO}} ^2$
are introduced.}}\label{tab:Chi_T_min}
\begin{center}
\begin{tabular}{cccc}
\hline\hline
$\beta$ & $\omega_1$ & $\Omega_{\rm{m0}}$ & $\chi_{\rm{BAO}} ^2$ \\
\hline
$-0.243$ & $-1.053$ & $0.272$   & $16 \times 10^{-4}$, \\
\end{tabular}
\begin{tabular}{cccc}
\hline\hline
 $\chi_{\rm{CMB}} ^2$ & $\chi_{\rm{OHD}} ^2$ &  $\chi_{\rm{SNe}} ^2$ & $(\chi_{\rm T} ^2)_{\rm{min}}$\\
\hline
 $12 \times 10^{-5}$ & $16.23$ &   $542.75$ & $558.98$\\
\hline\hline
\end{tabular}
\end{center}
\end{table}
\renewcommand{\arraystretch}{1.1}
\begin{table}
\caption{\small{This table is related to minimum point of
$\chi_{\rm T} ^2 = \chi_{\rm{SNe}} ^2 +  \chi_{\rm{OHD}} ^2 + 3 \chi_{\rm{CMB}} ^2 + 3\chi_{\rm{BAO}}^2$.}}\label{tab:Chi_T_min_w}
\begin{center}
\begin{tabular}{cccc}
\hline\hline
$\beta$ & $\omega_1$ & $\Omega_{\rm{m0}}$ & $\chi_{\rm{BAO}} ^2$\\
\hline
$-0.239$ & $-1.051$ & $0.272$   & $5 \times 10^{-4}$ ,\\
\end{tabular}
\begin{tabular}{cccc}
\hline\hline
$\chi_{\rm{CMB}} ^2$ & $\chi_{\rm{OHD}} ^2$ &  $\chi_{\rm{SNe}} ^2$ & $(\chi_{\rm T} ^2)_{\rm{min}}$\\
\hline
 $8 \times 10^{-10}$ & $16.23$ &   $542.75$ & $558.98$\\
\hline\hline
\end{tabular}
\end{center}
\end{table}


\textbf{\begin{table}
\caption{\small{In this table the quantities which maximize the relative probability functions  $\mathcal{L}(\Omega_{\rm{m0}})$, $\mathcal{L}(\omega_1)$ and $\mathcal{L}(\beta)$  using confidence levels $\sigma_1 = 68.3\%$ and    $\sigma_2 = 95.4\%$  are calculated. The data sets are includes of $SNeIa$, $CMB$, $BAO$ and $OHD$ in which the weight of  $\chi_{CMB}^2$ and $\chi_{BAO}^2$
in $\chi_{Total}^2$ function is the coefficient $3$.}}\label{tab:Moste_P_marginalized}
\begin{center}
\begin{tabular}{cccccc}
\hline
$\sigma_2 ^-$  & $\sigma_2 ^+ $ &  $\sigma_1 ^-$ & $\sigma_1 ^+$ & $(\mathcal{L})_{\rm{max}}$ &  x \\
\hline
0.02 & 0.02 &  0.01 & 0.01& 0.272 &  $\Omega_{m0}$ \\
\hline
0.16 & 0.156 &  0.08& 0.08 & -1.04 &  $\omega_1$ \\
\hline
0.44 & 0.59&  0.23 & 0.27 & -0.24 &  $\beta$\\
\hline
\end{tabular}
\end{center}
\end{table}}

\end{document}